\documentclass[prc,twocolumn,superscriptaddress,nofootinbib,showkeys,floatfix]{revtex4-1}
\usepackage{epsfig}
\usepackage{graphicx}
\usepackage{palatino}
\usepackage[english]{babel}
\usepackage{hyphenat}
\usepackage{amsmath}
\usepackage{amssymb}
\usepackage{mathtools}
\usepackage{mathrsfs}
\usepackage{slashed}
\usepackage{epstopdf}
\usepackage{xcolor}
\usepackage{braket}
\usepackage{booktabs}
\definecolor{lcolor}{rgb}{0.,0.0,0.}
\definecolor{citcolor}{rgb}{0,0.,0.5}
\usepackage[breaklinks,colorlinks,urlcolor=blue,citecolor=blue,linkcolor=blue]{hyperref}
\usepackage{multirow}
\usepackage{ltablex}
\usepackage{soul}

\def\bfxi{\mbox{\boldmath$\xi$}}
\newcommand{\beq}{\begin{eqnarray}}
\newcommand{\eeq}{\end{eqnarray}}

\newcommand{\bem}{\begin{multline}}
\newcommand{\eem}{\end{multline}}
\newcommand{\beg}{\begin{gather}}
\newcommand{\eeg}{\end{gather}}

\newcommand{\nn}{\nonumber\\}

\newcommand{\ben}{\begin{eqnarray*}}
\newcommand{\een}{\end{eqnarray*}}

\newcommand{\eqn}[1]{Eq.~\eqref{#1}}

\def\cG{{\cal G}}
\def\cP{{\cal P}}

\def\cO{{\cal O}}

\newcommand{\secn}[1]{Section~1}
\newcommand{\appn}[1]{Appendix~1}

\long\def\comment#1{ }

\def\Tr{\text{Tr}}

\def\and{\quad\text{and}\quad}

\newcommand{\rmd}{{\rm d}}

\newcommand{\rme}{{\rm e}}

\def\bell{{\boldsymbol \ell}}
\def\q{{\boldsymbol q}}
\def\0{{\boldsymbol 0}}
\def\p{{\boldsymbol p}}

\def\k{{\boldsymbol k}}

\def\0{{\boldsymbol 0}}
\def\x{{\boldsymbol x}}
\def\y{{\boldsymbol y}}
\def\X{{\boldsymbol X}}

\def\r{{\boldsymbol r}}
\def\s{{\boldsymbol s}}

\def\b{{\boldsymbol b}}
\def\Q{{\boldsymbol Q}}

\def\M{{\boldsymbol M}}
\def\K{{\boldsymbol K}}

\def\T{{\boldsymbol T}}

\newcommand{\del}{\partial}

\begin{document}

\title{Quantum to classical parton dynamics in QCD media}

\author{João Barata}
\email[]{jlourenco@bnl.gov}
\affiliation{Physics Department, Brookhaven National Laboratory, Upton, NY 11973, USA}
\author{Jean-Paul Blaizot}
\email[]{jean-paul.blaizot@ipht.fr}
\affiliation{Institut de Physique Théorique, Université Paris Saclay, CEA, CNRS, F-91191 Gif-sur-Yvette, France}
\author{Yacine Mehtar-Tani}
\email[]{mehtartani@bnl.gov}
\affiliation{Physics Department, Brookhaven National Laboratory, Upton, NY 11973, USA}
\affiliation{RIKEN BNL Research Center, Brookhaven National Laboratory, Upton, NY 11973, USA}

\begin{abstract}
We study the time evolution of the density matrix of a high energy quark propagating in a dense QCD medium  where it undergoes elastic collisions  (radiation is ignored in the present study). The medium is modeled as a stochastic color field with a Gaussian correlation function. This allows us to eliminate the medium degrees of freedom and obtain a simple master equation for the evolution of the reduced density matrix of the high energy quark, making use of approximations that are familiar in the description of open quantum systems.  
This master equation is solved analytically, and we demonstrate that its solution can be reconstructed from a simple Langevin equation.  At late times,  one finds that only the color singlet component of the density matrix survives the quark's propagation through the medium.  The off-diagonal elements of the density matrix are suppressed  successively in transverse position space and in momentum space, and become independent of the details of the initial condition. This behavior is reflected in the corresponding von Neumann entropy, whose growth at late time is related to the increase of the classical phase space explored by the high energy quark in its motion through the medium. The interpretation of the Wigner transform as a classical distribution is further supported by the fact that the associated classical entropy coincides at late time with the von Neumann entropy.  
\end{abstract} 


\maketitle

\section{Introduction}
\label{sec:intro}

Jets provide one of the best probes to explore the quark gluon plasma (QGP) produced in the aftermath of ultra-relativistic heavy ion collisions~\cite{Apolinario:2022vzg,Busza:2018rrf}. The medium modifications imprinted  into the jet structure, colloquially referred to as jet quenching, generally result in the emission of extra soft radiation outside of the jet cone and to the broadening of the jet structure~\cite{Blaizot:2015lma}. A detailed understanding of such medium induced effects is one of the major drives behind jet quenching theory.

Although the complete description of in-medium jet evolution requires a quantum treatment~\cite{Arnold:2002zm,Zakharov:1996fv,Baier:1996sk,Sievert:2018imd}, it turns out that, for some observables,  medium induced modifications admit a purely classical description, see e.g.~\cite{Blaizot:2012fh,Blaizot:2013vha}. The prototypical example is, at leading order and under the usual approximations, the effect of medium induced momentum broadening on the final particle distribution. This observable can be described as a convolution of the vacuum particle spectrum and a classical distribution encapsulating the medium induced modifications.\footnote{This is not generally true, as higher order corrections and other medium effects cannot be described solely using a classical approach~\cite{Barata:2022utc,Blaizot:2014bha,Sadofyev:2022hhw,Caucal:2022fhc, Liou:2013qya, Ghiglieri:2022gyv,Barata:2022krd, Andres:2022ndd, Barata:2022wim,Li:2020uhl}.
}   

More generally, one can capture the full jet dynamics by constructing the associated reduced density matrix, obtained by integrating out the medium degrees of freedom. The  particle distribution measured in the final state, from which one can study momentum broadening, is related to the diagonal elements in momentum space of this reduced density matrix, and it mainly reflects the classical aspects of the jet evolution. As a consequence, by constraining oneself to study only such observables, one leaves unexplored some of the quantum features of the in-medium jet evolution. From a theoretical point of view, it is desirable to extend the jet quenching formalism to track the jet's full quantum evolution and better identify the limits of a classical description. In this work, we make a first step in this direction by constructing the reduced density matrix of a single energetic quark in interaction with a dense QCD medium. The obtained  density matrix can be identified with that of a jet in the absence of radiation. This   simplified setting  allows us to follow the full evolution of the jet in the medium within an exact analytical treatment. 

It is interesting here to note the analogy between this study and that of the  propagation of heavy quarkonia in a QGP.  The  jets produced in heavy ion collisions  undergo substantial final state interactions with the QGP in which they begin  their evolution, and, as is the case for quarkonia,  it is fruitful to view them  as  open quantum systems,  with a non-unitary evolution \cite{Blaizot:2017ypk,Blaizot:2018oev,Akamatsu:2020ypb,Vaidya:2020cyi,Yao:2021lus}. The  master equations for the corresponding reduced density matrices share many common properties, and in particular they include the same collisional decoherence effects. Investigating the exact mechanisms by which QCD jets lose quantum and color coherence, as well as their energy, to the plasma, is not only fundamental for the understanding of out-of-equilibrium QCD dynamics but is also crucial to best exploit jets as probes of the QGP in heavy ion experiments. 

The focus of the present study concerns  the emergence of the classical evolution of the (simplified) jet. 
As we shall see,   the associated reduced density matrix  becomes asymptotically diagonal successively in position and momentum representations, suggesting indeed an interpretation  in terms of a classical phase space distribution. This interpretation is supported by the calculation of the
von Neumann entropy which, at sufficiently late times, takes the form 
\beq\label{eq:master_entropy}
  S_{\rm vN}(t) \approx  \log \Phi(t) \,.
\eeq
Here $\Phi(t)\propto \langle \K^2 \rangle_t \langle \b^2\rangle_t$ may be interpreted as a measure of  the two dimensional phase space of the quark in terms of its average squared total transverse momentum $\K$ and its transverse location $\b$ at time $t$. We find, in particular, that the emergent diagonalization of the quark density matrix  is accompanied by  a fast increase of the entropy, driven by  the fast growth of $ \langle \b^2\rangle_t$ at late times. The earlier suppression of the off-diagonal matrix element of the coordinate space density matrix is also accompanied by an entropy growth, driven by the increase of $\langle \K^2 \rangle_t$ and related to collisional decoherence. 

The present work is organized as follows. In section~\ref{sec:RDM} we introduce the quark reduced density matrix in the medium and obtain its evolution equation. We discuss the exact solutions to this equation in section~\ref{sec:color-random} in the so-called harmonic approximation. Within the setting considered and the approximations made, the problem belongs to a  class of problems that have been thoroughly studied in the literature of open quantum systems \cite{breuer2002theory}. Finally,  we compute the associated von Neumann entropy in section~\ref{sec:entropy}. In section~\ref{sec:sum} we summarize our findings. Appendix~\ref{app:dev_13_14} provides some technical details on the derivation of the evolution equations used in the main body of the work. Appendix~\ref{se:vNentropy} presents the calculation of the von Neumann entropy.

\section{The master equation for the reduced density matrix}
\label{sec:RDM}
\begin{figure}[]
  \centering
  \includegraphics[width=.3\textwidth]{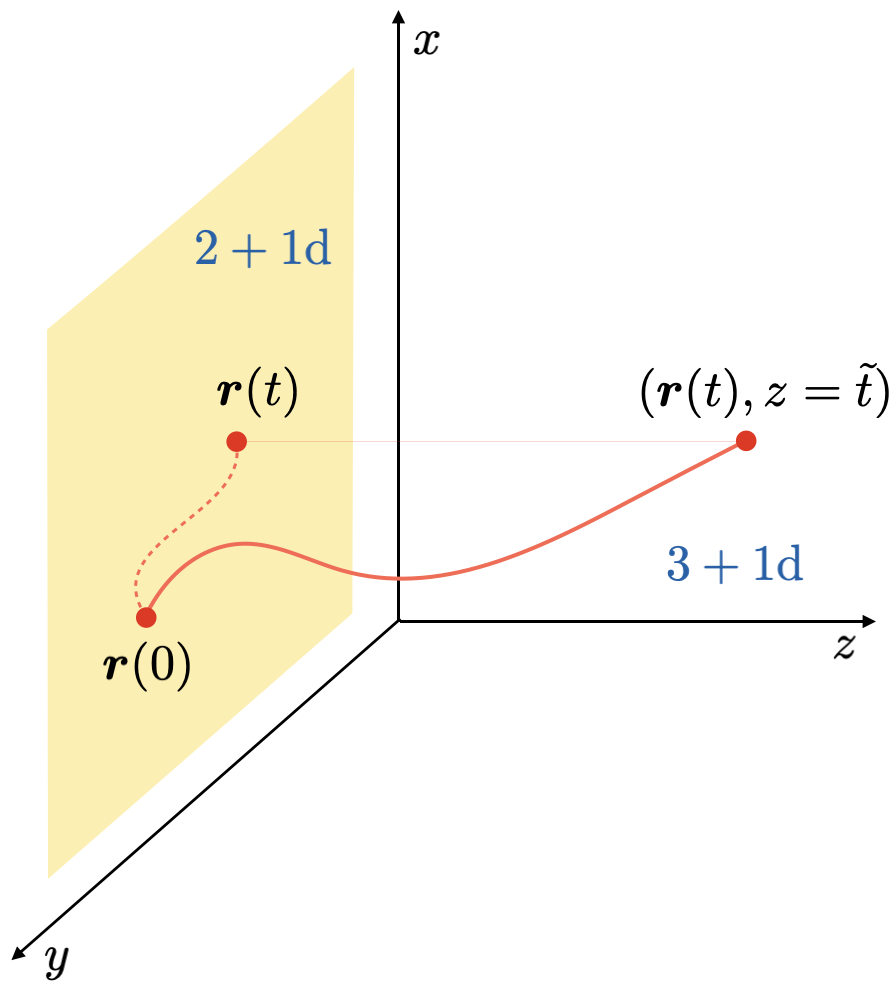}
  \caption{Depiction of the quark's propagation in three spatial dimension (at amplitude level), denoted by the euclidean coordinates $(x, y, z)$, along the positive $z$ axis. Since the parton is moving at the speed of light, its dynamics are constrained to the hyperplane $\tilde t-{z}=0$, with $\tilde t$ the usual time coordinate. As a result, the real time evolution of the quark state can be studied as the motion of an effective non-relativistic particle in 2+1 dimensions, evolving according to Eq.~\eqref{SchroG}.}
  \label{fig:doodle_draw}
\end{figure}

We consider the evolution of a highly energetic quark traveling in a dense medium of color charges at the speed of light in the positive $z$ direction.  This energetic parton couples to the long wavelength modes of the fluctuating gauge potential $A^{a,\,\mu}(\r,t)$ describing the medium. Here $t\equiv x^+$ denotes the light-cone time\footnote{In this work, we  use light-cone coordinates such that the light-cone time is defined as $t\equiv (\tilde t+z)/{\sqrt{2}} = \tilde t\sqrt{2}$, with $\tilde t$ the usual time coordinate and $z$ the longitudinal coordinate, see Fig.~\ref{fig:doodle_draw}. The last equality follows from the fact that we work in the frame where the quark is infinitely boosted along the positive $z$ direction. Also, in this frame, any dependence on the coordinate $x^-\equiv(\tilde t+z)/{\sqrt{2}} $ vanishes. For more details see e.g.~\cite{Blaizot:2015lma}.}  and $\r$ the coordinates of the fast quark transverse to the $z$ direction. Because the light-cone energy of the incident quark, $E\equiv p^+$, is taken to be much larger than that of the medium constituents, the essential dynamics is localized in the transverse plane, with the quark coupled to the minus component of the background field, $A^-(\r,t)$. The quark spin is conserved throughout the evolution.  The motion of the quark can be shown to be determined by the two-dimensional  Schr{\"o}dinger equation~\cite{Mehtar-Tani:2006vpj}
\begin{align}\label{SchroG}
\left[i\partial_t + \frac{\del^2_\perp}{2 E}+gA(\r,t) \right] \psi(\r,t)=0 \, ,
\end{align}
where $E$ plays the role of a mass. In this equation,  $A\equiv A^{a,\,  -} t^a$, with $t^a$ the generators of SU($N_c$) in the fundamental representation  for $N_c$ colors,  $\partial_\perp^2$ denotes the Laplacian operator in transverse space, and $\psi(\r,t)$ is the quark's wavefunction.      Thus, for highly energetic partons, the 3+1d evolution is dimensionally reduced to 2+1d, with the non trivial dynamics being constrained to the transverse plane at fixed light-cone energy $E$, see Fig.~\ref{fig:doodle_draw}.

The reduced density matrix $\rho$ of this energetic quark is obtained after taking the partial trace of the full density matrix $\rho[A]$ over the medium's degrees of freedom, i.e. over the gauge potential $A$ that enters Eq.~(\ref{SchroG}). In the present work, the trace over these gauge field configurations is  approximated by a statistical average, that is, the reduced density matrix  is given by
\begin{align}\label{rhoArho}
\rho \equiv {\rm tr}_{\!_A} \!\left( \rho[A] \right)= \Big\langle  |\psi_A(t) \rangle \langle \psi_A(t) |  \Big\rangle_{\!A}\,.
\end{align}
where $\ket{\psi_A}$ denotes the solution of the Schr{\"o}dinger Eq.~\eqref{SchroG} for a given field configuration $A(\r,t)$, and the field average is performed over a gaussian distribution whose two-point function is given by  
\begin{align}\label{eq:2-pt-correlator}
 g^2\Big \langle A^{a} (\q,t)A^{\dagger b} ( \q',t') \Big \rangle_{\!A} &= \delta^{ab } \delta(t-t')\nn 
 &\times (2\pi)^2 \delta^{(2)}(\q-\q')\, \gamma(\q) \, .
\end{align}
Recall that the  function $\delta(t-t')$  in Eq.~(\ref{eq:2-pt-correlator}) stands for $\delta(x^+-x^{\prime+})$. The medium correlations have a finite extent in $x^+$, of the order of the inverse of the Debye mass for a thermalized plasma. When the quark energy is large, because of time dilation, the quarks sees these correlations as if they were \textit{instantaneous} \cite{Blaizot:2012fh}. Similar considerations are involved in order to obtain Markovian equations for the  heavy quark problem  \cite{Blaizot:2017ypk}.

In writing Eq.~(\ref{rhoArho}), we have assumed that, prior to the averaging over the gauge field, the density matrix is that of a pure state, i.e., $\rho$ is the projector on the solution $\ket{\psi_A(t)}$ of  Eq.~\eqref{SchroG} for a given $A(\r,t)$.
The quantity $\gamma$ encodes the form of the effective interaction between the quark and the medium. More precisely, the average over the gauge field fluctuations is to be seen as a simple and efficient way to take into account the collisions of the high energy quark with the medium constituents, and  $\gamma$ is thus related to high energy limit of the in-medium elastic scattering rate, see e.g. Refs.~\cite{Blaizot:2015lma,Barata:2020rdn}. In the limit of high momentum transfer it scales as $\gamma(\q) \approx g^4 n /\q^{4}$, where $n$ is the density of color charges in the medium and $g$ is the strong coupling constant.

We already emphasized  in the introduction 
the analogy between the treatment of heavy quarks propagating in a quark-gluon plasma~\cite{Blaizot:2017ypk}, and that presented here for the motion of a jet in the transverse plane, with the energy jet $E$ playing the role of the large heavy quark mass. In both cases, it is convenient to treat the effect of collisions as an averaging  over a fluctuating background field\footnote{In the formulation of \cite{Blaizot:2017ypk}, the averaging over the gauge field yields an imaginary potential.}. We should also note the analogy with the strategy used in \cite{Blaizot:1997kw} to treat the effect of soft photons or gluons on the propagation of a hard fermion. There,  a corresponding averaging is made with an approximate quantum action.

We are now equipped to compute the quark density matrix. As a matrix in color space, it can be decomposed into singlet ($\rho_{\rm s}$) and octet ($\rho_{\rm o}$) components:
\begin{align}\label{eq:os_decomp}
 \rho(t)&\equiv  \rho_{\rm s} + t^a\rho^a_{\rm o} \nn 
 &= \frac{1}{N_c}\Tr_c (\rho) + 2\,  t^a \Tr_c(t^a\rho) \, ,
\end{align} 
where $\Tr_c$ denotes the trace over the fundamental color indices. Both $\rho_{\rm s}$ and $\rho^a_{\rm o}$, with $a=1, \, 2, \, \cdots, \, N_c^2-1$, are operators in transverse space. Their matrix elements in the coordinate representation read 
 \begin{align}\label{rhorrr}
\bra{\r}\rho_{\rm s,o}(t)\ket{\bar \r}=\bra{\b+\x/2}\rho_{\rm s,o}(t)\ket{\b-\x/2} \, ,
\end{align} 
where 
\begin{align}
\b\equiv \frac{\r+\bar\r}{2} \,,\quad  \x\equiv \r-\bar\r \,.
\end{align}
We shall denote by $\rho(\b,\r,t)$  the function given in \eqn{rhorrr} (with $\rho$ being either $\rho_{\rm s}$ or $\rho_{\rm o}^a$) and with a slight abuse of notation, we shall denote by $\rho(\bell,\K,t)$ the momentum space matrix element $\bra{\k}\rho\ket{\bar \k}$, where 
\begin{align}
\K\equiv \frac{\k+\bar\k}{2}\,,\quad\bell\equiv\k-\bar\k\,,
\end{align}
 are the variables conjugate respectively to $\x$ and $\b$ in the Fourier transform:\footnote{In this work we denote two dimensional momentum integrals as $\int d^2\q/(2\pi)^2\equiv \int_\q$, while for position integrals we use $\int d^2\x \equiv \int_\x$.}
\begin{align}\label{eq:rho_lk}
\rho(\bell,\K,t)\equiv \int_{\b,\x}\rme^{-i\bell\cdot\b}\rme^{-i\K\cdot\x} \, \rho(\b,\x,t).
\end{align} 
We shall also use mixed representations, such as the Wigner transform $\rho_{_W}\!(\b,\K,t)$\footnote{Note the abuse of notation: we denote by the same symbol $\rho$ different functions. The arguments of the function should suffice to lift the possible ambiguities. Note that the first argument refers to either $\b$ or $\bell$, and the second to either $\x$ or $\K$. Because of it special role in the present discussion we singularize the Wigner transform with the specific notation $\rho_{_W}$.} which is the Fourier transform of Eq.~\eqref{rhorrr} with respect to $\x$, and similarly for $\rho(\bell,\x,t)$.
The Wigner transform $\rho_{_W}\!(\b,\K,t)$ is real since $\rho=\rho^\dagger$ is hermitian, and normalized to unity (${\rm Tr}\rho=1$). It allows a natural connection with the classical regime, acquiring there the interpretation of a phase space distribution. Note however that, in contrast to a classical phase-space distribution, $\rho_{_W} $ need not be positive definite.\footnote{The Wigner function associated to the particular initial condition considered later in this paper is positive definite, as follows from Hudson's theorem \cite{hudson1974wigner}.} Loosely speaking $\rho_{_W}\!(\b,\K)$  gives the probability to find the quark at the location $\b$ with the momentum $\K$. This interpretation is supported by the fact that the integral over $\K$ of $\rho(\b,\K)$ is the probability to find the quark at  position $\b$,  $\rho(\b)\equiv\int_\K \rho_{_W}\!(\b,\K)=\bra{\b}\rho\ket{\b}=\rho(\x=0,\b)$\footnote{In this paper we shall often refer to $\rho(\b)$ as the density.}, while the integration over $\b$ is the transverse momentum distribution, ${\cal P}(\K)\equiv \int_\b \rho_{_W}\!(\b,\K)=\bra{\K}\rho\ket{\K}=\rho(\bell=0,\K)$. The latter is the quantity that usually appears in jet quenching observables \cite{Blaizot:2015lma}, since only the momenta of the final particles are measured, as emphasized in the introduction. 

\begin{figure}
  \centering
  \includegraphics[width=.3\textwidth]{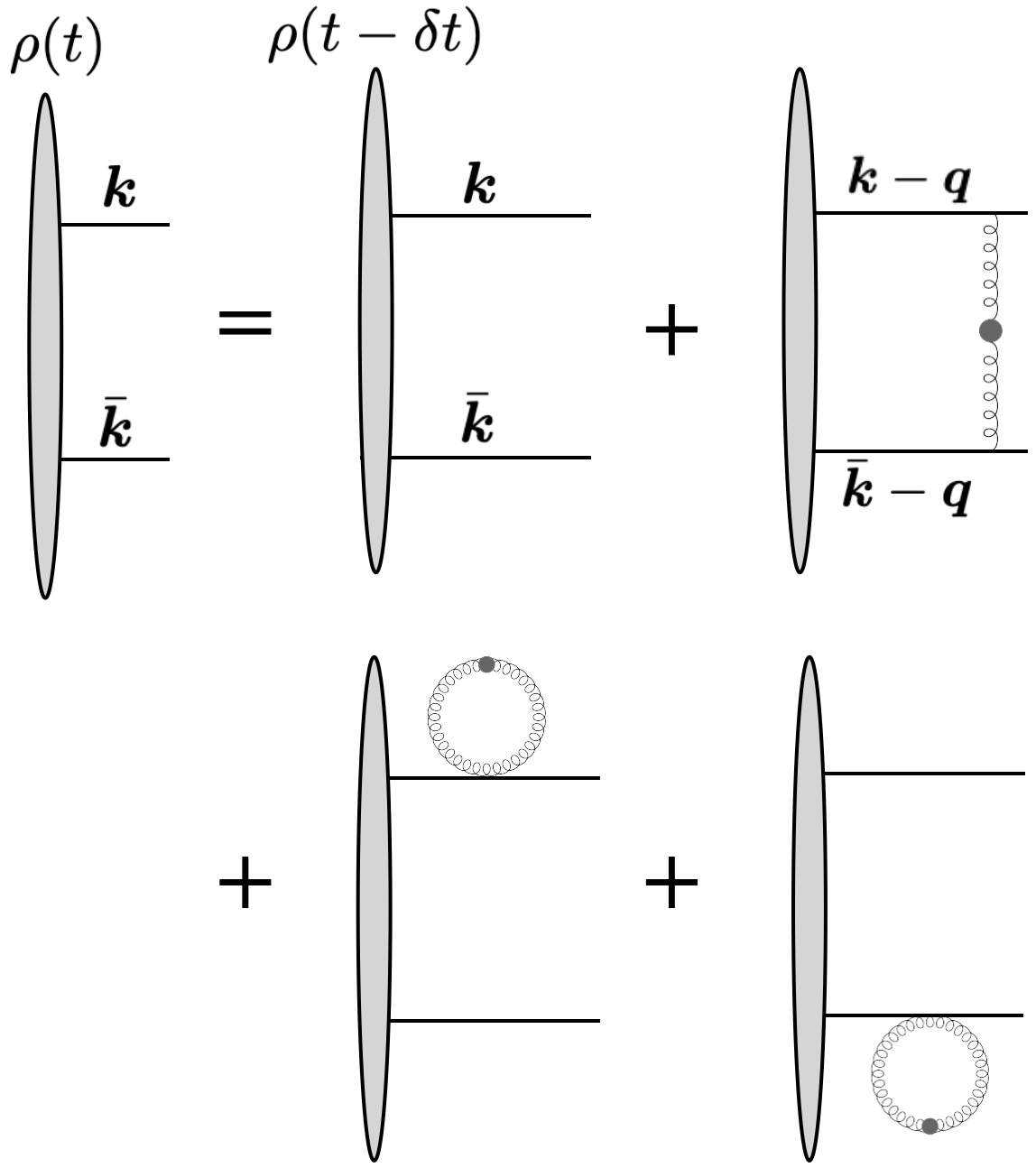}
  \caption{Graphs contributing to the transverse space evolution of the quark density matrix in an infinitesimal time step $\delta t$. Tadpole diagrams denote zero momentum exchange interactions, while the single scattering diagram denotes real momentum transfer with the medium.}
  \label{fig:master_draw}
\end{figure}

With these definitions, we can derive  the equations of motion for $\rho_{\rm s,o}(\bell,\K)\equiv \langle \k | \rho_{\rm s}(t)| \bar \k\rangle$ (see Appendix~\ref{app:dev_13_14} for a derivation)
 \begin{align}\label{eq:rho_s-mom}
  \langle \k | \rho_{\rm s}(t)| \bar \k\rangle&=C_F  \int_{\q}  \int_0^t dt' \, e^{i\frac{(\k^2-\bar \k^2)}{2E}(t-t')} \nn 
   &\times  \gamma(\q)\left[   \langle \k-\q | \rho_{\rm s}(t')| \bar \k-\q\rangle     -  \langle \k | \rho_{\rm s}(t')| \bar \k\rangle   \right]\,,
\end{align}
and 
 \begin{align}\label{eq:rho_o-mom}
  &\langle \k | \rho_{\rm o}(t)| \bar \k\rangle=C_F \int_{\q}  \int_0^t dt' \, e^{i\frac{(\k^2-\bar \k^2)}{2E}(t-t')} \nn 
   &\times  \gamma(\q)\left[   \langle \k-\q | \rho_{\rm o}(t')| \bar \k-\q\rangle     +  \frac{1}{2N_c  C_F}  \langle \k | \rho_{\rm o}(t')| \bar \k\rangle   \right]\,,
\end{align}
for the singlet and octet components, respectively. The two terms in the r.h.s. of Eqs.~(\ref{eq:rho_s-mom}, \ref{eq:rho_o-mom}) are illustrated in Fig.~\ref{fig:master_draw}.  The equations iterate the elementary processes depicted in Fig.~\ref{fig:master_draw} to all orders in the number of field insertions. Remarkably, the evolution equations for $\rho_{\rm s}$ and $\rho_{\rm o}$ are decoupled. This can be traced back to the fact that the two-point function in Eq.~(\ref{eq:2-pt-correlator}) is diagonal in color space and thus does not change the color structure of the density matrix.  In the graphical representation of Fig.~\ref{fig:master_draw} this color structure is that of the fictitious color dipole formed by the two horizontal lines that carry the color indices of the density matrix represented by the grey blob. That the instantaneous one-gluon exchange does not change this color structure can then be understood by  calling on the identities $t^at^a =  C_F $ and $t^at^bt^a = -\frac{1}{2N_c} t^b $, in the singlet and octet cases, respectively. 

The derivation of Eq.~(\ref{eq:rho_s-mom}) presented in Appendix~\ref{app:dev_13_14} relies on the formalism developed in \cite{Blaizot:2012fh}. An alternative derivation follows that used for heavy quarks in \cite{Blaizot:2017ypk}. Equation~(\ref{eq:rho_s-mom}) is in fact a simplified version of the Lindblad equation derived in \cite{Blaizot:2017ypk}, obtained by ignoring there the dissipative terms. It reads, in the position representation
\begin{align}\label{eq:rho_s_position1}
    \frac{\partial}{\del t} \langle  \r| \rho_{\rm s,o}(t)|\bar \r  \rangle  &=  -\frac{i}{2E}  \left(\frac{\del^2}{\del \bar\r^2}-\frac{\del^2}{\del \r^2}\right) \, \langle  \r| \rho_{\rm s,o}(t)|\bar \r  \rangle \nn
    &\quad -\Gamma_{\rm s,o} (\bar \r-\r)\, \langle  \r| \rho_{\rm s}(t)|\bar \r \rangle    \, .
\end{align}
The first term on the right-hand side represents unitary evolution,  which reduces here to free motion. It could be written $\bra{\r} \left[ H_0,\rho_{\rm s,o}\right]\ket{\bar\r}$, with $H_0=-(1/2E)\del_\r^2$ the transverse kinetic energy. The second term proportional to $\Gamma$ accounts for non-unitary evolution and, as we shall see, is directly connected to decoherence in coordinate space.\footnote{Eq.~(\ref{eq:rho_s_position1})  is  similar to that introduced in \cite{Joos:1984uk} to study  collisional decoherence. } In Eq.~(\ref{eq:rho_s_position1}), 
 $\Gamma_{\rm s,o} $ stand for the singlet and octet damping rates ~\cite{Blaizot:2013vha}. They are related to $\gamma(\q)$ by the following identities
 \begin{align}\label{eq:rho_o}
  \Gamma_{\rm s}(\x) &=C_F  \int_\q \left(1-e^{i \q \cdot \x}\right)\gamma(\q) \, , \nn 
  \Gamma_{\rm o}(\x) &=  \int_\q \left(C_F+\frac{1}{2N_c }e^{i \q \cdot \x}\right)\gamma(\q) \, .
\end{align}
with $C_F=(N_c^2-1)/(2N_c)$.

In order to solve Eqs.~(\ref{eq:rho_s-mom}) and (\ref{eq:rho_o-mom}) or equivalently Eqs.~(\ref{eq:rho_s_position1}),   it is convenient to work in the mixed representation $(\bell,\x)$, where these equations take the form:
\begin{align}\label{eq:rho_s}
    \partial_t \rho_{\rm s,o}(\bell,\x,t) &=  -\left[  \frac{\bell \cdot \partial_\x}{E} 
  + \Gamma_{\rm s,o}(\x)\right] \rho_{\rm s,o}(\bell,\x,t)  \, ,
\end{align}
The $(\bell,\x)$ mixed representation is convenient because the interaction term is local in $\x$ while the partial Fourier transform diagonalizes part of the kinetic energy. Eqs.~\eqref{eq:rho_s} can be formally solved by acting on them with the translation operator $\exp\left[(\bell \cdot \partial_\x) t/E \right]$, leading to (see also Eq.~\eqref{eq:klop}) 
\begin{align}\label{eq:help_sol}
  \partial_t \rho_{\rm s,o}(\bell,\y(t),t) &=  - \Gamma_{\rm s,o}(\y(t)) \rho_{\rm s,o}(\bell,\y(t),t)  \, ,
\end{align}
where $\y(t)=\x+ \frac{\bell\,}{E}   t $ and we used the property 
\begin{align}
e^{\frac{\bell \cdot \partial_\x}{E} t } f(\bell,\r,t) =  f\left(\bell,\x+ \frac{\bell}{E} t,t\right) \, ,
\end{align}
with $f$  a test function. From  Eq.~\eqref{eq:help_sol} one readily obtains
  \begin{align}\label{eq:master_Wigner} 
     \rho_{\rm s,o}(\bell,\x,t) &= \, \rho_{\rm s,o}^{(0)}\left(\bell,\X(t)\right) 
  \rme^{- \int_0^t d u \, \Gamma_{\rm s,o}\left(\X(u)\right)} \, ,
\end{align}
where now $\X(t)=\x - \frac{\bell }{E}   t $ and $\rho_{\rm s,o}^{(0)}\left(\bell,\x\right) =\rho_{\rm s,o}(\bell,\x,0)$ is the initial condition.  

There are two limiting cases where the solution is easily interpreted. The first case corresponds to the absence of interactions with the medium. Then, $ \rho_{\rm s,o}(\bell,\x,t) = \rho_{\rm s,o}^{(0)}\left(\bell,\X(t)\right) $  describes free streaming, where the relative coordinate $\X(t)$ evolves in time  at constant velocity $\bell/E$. The Wigner transform  corresponding  to this solution is $\rho_{_W}(\b-t\K/E,\K,t=0)$, showing in particular that, as expected,  the momentum distribution is not affected by the free motion, i.e. $\int_\b \rho_{_W}(\b,\K,t) = \int_\b  \rho_{_W}(\b-t\K/E,\K,0) = {\cal P}(\K,0)$. 

The other simple solution corresponds to the case where the drift term can be ignored, which occurs  either for $\bell=0$ or $E\to\infty$, i.e. the exact eikonal limit. This corresponds to the infinite mass limit in the heavy quark problem, a limit  where one expects classical features to emerge in the coordinate space description of the dynamics. In particular, in this limit $\x$ and $\b$ are independent of time and the solution reads
 \begin{align}\label{eq:master_Wigner2} 
     \rho_{\rm s,o}(\b,\x,t) &= \, \rho_{\rm s,o}^{(0)}\left(\b,\x\right) 
  \rme^{- t\,  \Gamma_{\rm s,o}\left(\x\right)} \, .
  \end{align}
This formula shows that the coordinate space density matrix is  damped at  rates $\Gamma_{\rm s,o}(\x)$ that depend solely on $\x$. These rates exhibit  interesting features in the limit of small $\x$. Indeed, when $\x\to 0$, $\Gamma_{\rm s}\to 0$ while  $\Gamma_{\rm o}\to -\frac{C_A}{2}\int_\q \gamma(\q)$, see Eq.~\eqref{eq:harm_calc} below. The first property, $\Gamma_{\rm s}\to 0$, is related to the phenomenon commonly referred to as color transparency~\cite{Brodsky:1988xz, Dutta:2012ii}.  Technically, it results from a  destructive interference  between the three diagrams in Fig.~\ref{fig:master_draw}. Physically, as argued earlier,  one may view the upper and lower lines of the diagrams in Fig.~\ref{fig:master_draw} as the two members of a fictitious color dipole propagating into the medium. A color singlet dipole of small size is seen as a neutral object by the medium, which results in the suppression of the interactions.

In the case of the octet  the aforementioned cancellation does not take place. In fact, in the limit of vanishing size, the octet behaves as a gluon, and the corresponding damping factor involves the so-called gluon damping rate~\cite{Braaten:1990it,Blaizot:1996az}. This suppression of the color octet component of the density matrix results in the  equilibration of colors: the density matrix of a quark initially in a given color state (generally containing both singlet and octet components) will, over a time scale $1/\Gamma_{\rm o}$,  turn into a singlet density matrix where all color states are equally populated~\cite{Blaizot:2017ypk}. A similar color equilibration was recently illustrated for the case of color antenna, with fixed kinematics, evolving in a dense medium~\cite{Zakharov:2018hfz}.

There is another effect of the damping $\Gamma_{\rm s}$: since it increases as $|\x|$ increases, it suppresses the non diagonal matrix elements of $\rho(\b,\x)$. This suppression of non diagonal matrix elements due to collisions is commonly referred to as decoherence. It will be discussed in more details later.

 An useful approximation for the rates $\Gamma_{\rm s,o}(\x)$  is  the so-called harmonic approximation~\cite{Zakharov:1996fv},  $\Gamma_{\rm s}(\x)  = \frac{1}{4}\hat q \,  \x^2 $, where $\hat q$ is the jet quenching parameter which will be specified shortly. This form correctly describes soft interactions between the jet and the medium and can be shown to be valid for most relevant models used in jet quenching~\cite{Barata:2020rdn,Djordjevic:2008iz,Barata:2020sav}. It corresponds to the approximation $\gamma= g^4 n/\q^4$ of $\gamma(\q)$, with which we get  
\begin{align}\label{eq:harm_calc}
  \Gamma_{\rm s}(\x) &\approx 4\pi \alpha_s^2 C_F n \log \left(\frac{Q^2}{m_D^2}\right)\frac{\x^2}{4} \equiv \frac{ \hat q}{4} \x^2 \, ,\nn 
  \Gamma_{\rm o}(\x) &\approx \frac{4\pi \alpha_s^2 C_A n}{ m_D^2}   \, ,
\end{align}
with $C_A=N_c$, and $\hat q$ is the jet quenching parameter in the fundamental representation. This is a logarithmically divergent quantity, and requires the introduction of an ultraviolet cut-off $Q$, which can be chosen as the characteristic saturation scale~\cite{Barata:2020sav},  as well as an infrared cut-off chosen to be the medium's Debye mass $m_D\sim g T$, with $T$ the medium's temperature.  Notice that since $\gamma\propto n \sim T^3$, then $\hat q \sim T^3$. We can also understand $\Gamma_{\rm o} \sim 1/\ell_{\rm fmp}\sim g^2 T$, as defining the inverse of a mean  free path $\ell_{\rm fmp}$ for gluons in the medium.

 Using the harmonic  approximation one can write Eq.~(\ref{eq:rho_s}) as the following (frictionless) Fokker-Planck equation for the Wigner function
 \begin{align}\label{eq:FPl}
 \del_t \rho_{_W}\!(\b,\K,t)=\left[-\frac{\K}{E}\frac{\del}{\del\b} +\frac{\hat q}{4}\frac{\del^2}{\del{\K^2}}\right]\rho_{_W}\!(\b,\K,t).\nn
\end{align}
 By integrating this equation over $\b$ one gets an equation for the momentum distribution
 \begin{align}\label{diffisionPK}
 \del_t \cP(\K,t)=\frac{\hat q}{4}\frac{\del^2}{\del\K^2}\cP(\K,t).
\end{align}
 This is a diffusion equation in momentum space, with $\hat q/4$ playing the role of diffusion constant. The solution to that equation with initial condition $\cP(\K,t=0)=(2\pi)^2\delta(\K)$ is simply 
 \begin{align}\label{calP}
   \cP(\K,t) =\frac{ 4\pi}{\hat q t} e^{-\frac{\K^2}{\hat q t}} \, .
\end{align}
As this formula illustrates, $\hat q t$ is the average transverse momentum acquired by the energetic quark as it traverses the medium during a time $t$, i.e. $\langle \K^2\rangle_t=\hat q t$.

As is well known, the dynamics encoded in the solution of the Fokker-Planck equation (Eq.~(\ref{eq:FPl})) can be equivalently obtained  by solving an associated Langevin equation, which in the present case takes the following simple form
\begin{align}\label{langevin}
E\frac{\rmd^2 \b}{\rmd t}=\bfxi(t),\quad \langle \xi_i(t_1)\xi_j(t_2)\rangle=\frac{\hat q}{2}\delta_{ij}\delta(t_1-t_2).
\end{align}
This equation describes the motion of a particle subjected to a random force $\bfxi(t)$, with the strength of the associated white noise controlled by the jet quenching parameter $\hat q$. This equation is easily solved.  As an illustration, consider  the momentum $\K=E\,\rmd \b/\rmd t$. By integrating Eq.~(\ref{langevin}) over time, one gets
\begin{align}\label{Knoise}
\K(t)=\K(t=0)+\int_0^t \rmd t' \bfxi(t'). 
\end{align}
This solution can be used to evaluate the expectation value $\langle\K^2(t)\rangle$ by averaging over the noise and the initial condition. Taking, for simplicity,  as initial condition $\K(t=0)=0$, one gets $\langle\K^2(t)\rangle=\hat q t$, in agreement with the result obtained earlier from the momentum distribution. 

In the following we shall analyze in detail various features of the solution of the equation of motion for the quark density matrix for a specific initial condition and using the harmonic approximation. Since, as we have argued above, the color octet component is damped when the energetic quark propagates in the medium, we now focus on the singlet component and study in more detail its time evolution. As we only consider singlet quantities, we drop from now on the subscript $\rm s$.

\section{Evolution of the singlet density matrix }
\label{sec:color-random}

To study the evolution of the singlet component, we will assume that initially the energetic quark is described by a gaussian wave packet $\psi_0(\k)$ centered around vanishing transverse momentum. The initial density matrix is then given by its momentum space matrix elements
\begin{align}\label{eq:init_cond}
 \bra{ \k}\rho(0)\ket{\bar \k}&=  \psi_0(\k)\psi_0^*(\bar\k)
 \hspace{0 cm}=   \frac{4\pi }{\mu^2}   \,   \rme^{-\frac{\k^2+\bar \k^2}{2\mu^2}}    \, ,
\end{align}
with $\mu $ a parameter characterizing the  extension  of the initial wave packet. The corresponding Wigner transform reads 
\begin{align}\label{Wig0}
 \rho_{_W}\!(\b,\K,0)=4 \rme^{-\mu^2 \b^2}\rme^{-\frac{ \K^2}{\mu^2}}\, .
\end{align}
This Wigner transform is akin to that of a coherent state, where the  particle is localized in both momentum space and coordinate space, with an accuracy specified by the single parameter $\mu$. 
Thus the dispersion in position and momentum are correlated and  when $\mu\to 0$, $ \rho(\K)\to (2\pi)^2 \delta(\K)$, while when $\mu\to \infty$, $ \rho(\b)\to\delta(\b)$. Note that, due to our specific choice of initial condition,  the Wigner transform is initially positive and remains so as it evolves in time, as we shall verify. 

In the absence of interactions, the initial  wave packet spreads freely and, as observed earlier, its Wigner transform evolves  as $\rho_{_W}\!(\b-(\K/E)t,\K,0)$. The momentum distribution remains unchanged, but the density $\rho(\b,t)$, obtained by integrating the Wigner transform over $\K$, spreads according to
 \begin{align}\label{rhofs}
 \rho(\b,t)=\frac{1}{\pi\langle \b^2\rangle_t^{(0)}} \rme^{-\frac{\b^2}{\langle \b^2\rangle_t^{(0)}}},\quad  \langle \b^2\rangle_t^{(0)}\equiv \frac{1}{\mu^2}\left( 1 +\frac{t^2}{t_0^2}\right)\, .
\end{align}
 Here the characteristic time scale 
 \begin{align}
  t_0\equiv \frac{E}{\mu^2} \, ,
 \end{align}
denotes the time at which the spreading of the wave packet starts to become significant; it corresponds to the time it takes the quark to cover, at velocity $\mu/E$, a distance equal to the spatial size $\sim 1/\mu$ of the initial wave packet. Note that the same result would have been obtained for a classical distribution function. In fact,  $\langle \b^2\rangle_t^{(0)}$, the time-dependent mean square radius of the  density, can be obtained via a simple argument that exploits the underlying classical dynamics. Indeed, the trajectory of the quark in the transverse plane is given by $\b(t)=\b(0)+\frac{\K}{E} t$, so that 
\begin{align}\label{b2nonoise}
\b^2(t)=\b^2(0)+2\frac{\K\cdot \b(0)}{E} t+\frac{\K^2}{E^2} t^2.
\end{align}
 By  averaging this expression over the initial conditions  encoded in Eq.~(\ref{Wig0}), i.e., $\langle\b^2(0)\rangle=1/\mu^2$, $\langle \K^2\rangle=\mu^2$, $\langle \K\rangle=0$,  one easily reproduces the result given in Eq.~(\ref{rhofs}).

When one allows for interactions, and employs the harmonic approximation in Eq.~(\ref{eq:master_Wigner}), one obtains the following expression for the density matrix in the mixed representation ($\bell,\x$):
\begin{align}\label{eq:master_Wigner_tt} 
     \rho(\bell,\x,t) &= \rho^{(0)}\left(\bell,\X(t)\right) \rme^{ -\frac{\hat q}{4} \int_0^t d u \left(\x(u)\right)^2}\nn
     &= \rme^{-\frac{1}{4} \left(a(t) \x^2 +b(t) \x\cdot\frac{\bell}{E} +c(t)\frac{\bell^2}{E^2}\right)}\, ,
    \end{align}
    where 
\begin{align}\label{eq:abc}
a(t)&=\mu^2+\hat q t\equiv\mu^2\left(1+\frac{t}{t_1}\right)\, , \nn
b(t)&=-2 \mu^2t-\hat q t^2 \equiv -2 \mu^2 \left(1+\frac{t}{2t_1}\right) t \, ,  \nn
c(t)&=\frac{E^2}{\mu^2}+{\mu^2t^2 } +\frac{\hat q t^3}{3}\equiv\frac{E^2}{\mu^2}\left[1+\frac{t^2}{t_0^2}+ \frac{1}{3}\frac{t^3}{t_2^3} \right]   \, .
\end{align}
The time scales $t_1$ and $t_2$ will be defined shortly (see Eqs.~(\ref{eq:t1def}) and (\ref{eq:t2def}) below).
The coefficients $a(t),b(t),c(t)$ have a simple physical interpretation which is best seen on the Wigner transform, given by 
\begin{align}
\rho_{_W}\!(\b,\K)=\frac{4}{D} \exp\left\{-\frac{1}{D}\left(a \,\b^2+ \frac{b}{E}\,  \b\cdot \K   +\frac{c}{E^2} \K^2 \right) \right\},\nn
\end{align}
with 
\begin{align}\label{coefabc}
a=\langle \K^2\rangle_t,\quad \frac{c}{E^2}=\langle \b^2\rangle_t,\quad \frac{b}{E}=-2\langle \b\cdot\K\rangle_t,
\end{align}
and 
\begin{align} 
D=\langle \b^2\rangle \langle \K^2\rangle-\langle \K\cdot\b\rangle=\frac{ ac}{E^2}-\frac{b}{2E}\,.
\end{align}
The  averages are taken with $\rho_{_W}$, for instance
\beq
\langle \K^2\rangle_t=\int_{\b,\K} \K^2 \rho_{_W}(\b,\K,t).
\eeq
Alternatively, the same expectation values can be obtained by solving the  Langevin equation (Eq.~(\ref{langevin})), and performing averages over both the noise (as in Eq.~(\ref{Knoise})) and the initial condition (as in Eq.~(\ref{b2nonoise})). Thus, to a large extent, the time evolution is  governed by the classical Langevin equation, quantum mechanics entering mainly in the averaging over the initial condition defined by the initial wave packet.

At this point, it is useful to make further contact with the formalism used in \cite{Blaizot:2012fh} (see also Appendix~\ref{app:dev_13_14}).  There, we introduced  a two-point function $S^{(2)} $ which plays a role similar to that of the Liouville operator that propagates the density matrix from the initial time to the final time (see Appendix~B1 in \cite{Blaizot:2012fh}) . We have indeed 
\beq\label{Liouville0}
\bra{ \r'} \rho(t) \ket{\bar\r'} =\int_{\r,\bar \r} S_{\r',\bar\r';\r,\bar\r}^{(2)}(t)\bra{\r}\rho(t=0)\ket{\bar \r},
\eeq
where (with a somewhat different notation than in \cite{Blaizot:2012fh})
\beq\label{UsingS2}
S_{\r',\bar\r';\r,\bar\r}^{(2)}(t)&=& \left(\frac{E}{2\pi t}   \right)^2 \exp\left\{  \frac{i E}{2 t}\left[ (\r'-\r)^2-(\bar \r'-\bar \r)^2\right]  \right\}\nonumber \\
&&\qquad \times \exp\left\{ -\frac{\hat q}{4} \int_0^t d u \left(\s(u)\right)^2\right\}.
\eeq
In this formula,   $\s(u)$ is a linear function of $u$ such that $\s(0)=\r-\bar \r$ and $\s(t)=\r'-\bar \r'$.
This expression of the density matrix is similar to  that in Eq.~(\ref{eq:master_Wigner}), and it also reflects the structure of Eq.~(\ref{eq:rho_s_position1}).  
 The first line in this expression is the product of a free propagator and a complex conjugate one, and represents free motion.   The second line is the exponential representing the effects of the collisions, as in Eq.~(\ref{eq:master_Wigner}). The two lines are not independent since the trajectory $\s(u)$ depends on the end points ($\r,\bar\r, \r',\bar \r'$).\footnote{Note that in \cite{Blaizot:2012fh}  no attempt was made to determine the density matrix. Rather one proceeded with  simplifications that were appropriate to get only the momentum distribution.} Given an initial density matrix of the form 
 of Eq.~(\ref{eq:init_cond}), viz.
\beq
\rho_0(\r,\bar\r)=\frac{\mu^2}{\pi}\exp\left\{-\frac{\mu^2}{2}(\r^2+\bar\r^2)  \right\},
\eeq
one can  explicitly calculate  the relevant Gaussian integrals involved in Eq.~(\ref{Liouville0}) and obtain for  $\rho(t)$ the following expression 
\beq\label{rhobx}
\rho(\b,\x,t) =\frac{1}{\pi} \frac{E^2}{c} \rme^{-\frac{E^2}{c}\b^2}\,\rme^{\left( - \frac{a}{4 }+\frac{b^2}{16c} \right)\x^2}\rme^{i\frac{bE}{2c} \x\cdot\b},  
\eeq
where  we have set $\b=(\r+\r')/2$ and $\x=\r'-\bar \r'$. It is easily verified that this is the same as Eq.~(\ref{eq:master_Wigner_tt}) after a Fourier transform of the variable $\b$.

We return now to the analysis of the physical content of  the density matrix. Its  momentum representation is obtained by a Fourier transform of $\rho_{_W}(\b,\K)$  with respect to $\b$. It reads 
\begin{align}\label{eq:rho_mom}
&\rho (\bell,\K)=\nn
&\frac{4\pi }{a} \exp\left\{- \frac{1}{4 a} \K^2  - \frac{1}{4 E^2 } \left(c-\frac{b^2}{4a}\right) \bell^2 - i \frac{b}{4 Ea }\bell\cdot \K \right\}.\nn
\end{align}
From this expression, the momentum distribution is obtained simply by setting $\bell=0$. It  is of the form displayed in Eq.~(\ref{calP}) with $\hat q t$ substituted by 
\begin{align}\label{kt2}
  \langle \k^2\rangle_t=\mu^2+\hat q t = \mu^2 \left(1+\frac{t}{t_1}\right)\, .
\end{align}
One can therefore identify two regimes, separated by the time scale 
\begin{align}\label{eq:t1def}
  t_1= \frac{\mu^2}{\hat q} \, .
\end{align}
When $t\ll t_1$, the collisions have little effect on the momentum distribution which retains its initial shape. When  $\hat q t\gg \mu^2$ or $t\gg t_1$,   collisions dominate and move the quark away from the region covered by the initial wave packet. 
 
Another time scale emerges when one considers the density $\rho(\b)$, obtained from $\rho(\b,\x)$ in Eq.~(\ref{rhobx}) by  setting $\x=0$. One gets 
\begin{align} \label{rhobt}
\rho(\b,t)= \frac{1}{\pi \langle \b^2\rangle_t} \, \rme^{  -\frac{\b^2}{\langle \b^2\rangle_t} } \, ,
\end{align}
where
\begin{align}\label{bt2}
  \langle \b^2\rangle_t=\frac{c(t)}{E^2}=\frac{1}{\mu^2}\left[1+\frac{t^2}{t_0^2}+ \frac{1}{3}\frac{t^3}{t_2^3} \right] \, ,
\end{align}
and the new time scale $t_2$ is given by (see Eqs.~\eqref{eq:abc} and \eqref{coefabc})
\begin{align}\label{eq:t2def}
   t_2^3\equiv \frac{E^2}{\hat q \mu^2} \, .
  \end{align}
We recognize in the first two contributions to $\langle \b^2\rangle_t$ (that are independent of $t_2$) the width of the Gaussian that corresponds to the collisionless evolution of the initial wave packet, i.e. $\langle \b^2\rangle_t^{(0)} $ in Eq.~(\ref{rhofs}): $ \langle \b^2\rangle_t =\langle \b^2\rangle_t^{(0)} +\hat q t^3/3 E^2$. The  initial spreading of the wave packet  is amplified by the momentum space diffusion, which eventually leads to a rapid spreading of the density, with 
\begin{align}\label{rhobt_latetimes}
 \rho(\b,t)\approx \frac{3 E^2}{\pi \hat q t^3} \exp\left\{-\frac{3E^2}{\hat q t^3} \b^2  \right\} \qquad  (t\gg t_2) \, ,
\end{align}
where we have used that, for $t\gg t_2$,  $\langle \b^2\rangle_t \simeq  {\hat q  t^3}/{3 E^2 }$. Note that, as was the case for the momentum distribution, in this late time regime, $\rho(\b,t)$ is independent of the scale $\mu$:  the memory of the initial condition is lost. Note however that the onset of this late time regime depends on $\mu$.

So far we have considered the diagonal components of the density matrix, either the momentum distribution or the spatial density. The evolutions of these diagonal components are driven by the natural spreading of the initial wave packet in coordinate space, together with diffusion in momentum space induced by collisions. As we have seen both phenomena are captured by a simple Langevin equation and they would not be different if the Fokker-Planck equation were solved for a classical phase-space distribution function, provided this distribution was initialized as in Eq.~(\ref{eq:init_cond}). We turn now to the off-diagonal components of the density matrix which encode quantum correlations. A priori,  the density matrix cannot be simultaneously diagonal in both momentum and position spaces, so we shall consider both representations in turn. We consider first the momentum space representation. To simplify the discussion, we focus on the case $\K=0$. We have
\begin{align}\label{eq:llop}
\rho(\bell,\K=0,t)= \frac{4\pi}{\mu^2(1+(t/t_1))} \exp\left\{  -\frac{\bell^2}{4\mu^2}  d(t)  \right\} \, ,
\end{align}
where 
\begin{align}
d(t)=1+\frac{1}{12}\left(  \frac{t}{t_2}\right)^3 \frac{t+4t_1}{t+t_1}\, . 
\end{align}
When $t\ll t_2$, $d(t)\ll 1$ and the $\bell^2$-distribution is only moderately affected by the collisions. In the opposite regime where $t\gg t_2$, the $t^3$  factor becomes dominant, and the $\mu$ dependence in the exponent cancels out, leaving us with
\begin{align}
  \rho(\bell,\K=0,t)&\approx \frac{4\pi}{\mu^2+\hat q t}\exp\left\{  -\frac{\bell^2\, \hat q t^3}{48 E^2}   \right\} \nn 
  &=\frac{4\pi}{\langle \k^2\rangle_t}\exp\left\{  -\frac{\bell^2 \langle \b^2\rangle_t }{16}   \right\} \, .
  \end{align}
The first factor accounts for the decrease of the momentum distribution ${\cal P}(\k=0,t)$ when $t\gtrsim t_1$. The other factor is a Gaussian distribution in $\bell=\k-\bar \k$, whose width decreases rapidly, as $1/t^3$, indicating that at late times the  density matrix becomes diagonal  in $\k$-space. The width of this distribution is the inverse of that in Eq.~(\ref{rhobt_latetimes}) characterizing the behavior of the density at late time.

A similar analysis can be performed in coordinate space. We have
\begin{align}
\rho(\b=0,\x,t)= \frac{E^2}{\pi c(t)} \exp\left\{-\frac{ E^2 a(t)}{4 \mu^2 c(t)} d(t) \x^2 \right\} \, .
\end{align}
At late times, this expression reduces to
\begin{align}\label{rhob0x2}
\rho(\b=0,\x,t)\approx \frac{1}{\pi \langle \b^2\rangle_t }\exp\left\{ -\frac{\langle \K^2\rangle_t \,\x^2}{4} \right\} \, ,
\end{align}
which shows that also in coordinate space the density matrix becomes diagonal. The width of the $\x$ distribution is now inversely proportional to  that of the momentum distribution, while the prefactor before the Gaussian represents the fast decrease of the density at $\b=0$ due to its spreading in the transverse plane. 

One important aspect of the calculation that leads to Eq.~(\ref{rhobx}), is that the collisions only affect the location $\b(t)$ of the particle, or its total momentum $\K=E\rmd \b/\rmd t$, but not the relative distance $\x$. This property emerges naturally in the path integral formulation (see the discussion after Eq.~(B6) in \cite{Blaizot:2012fh}). At the level of  Eq.~(\ref{eq:master_Wigner}) it is reflected in the explicit equation $\X(t)=\x-\frac{\bell}{E}t$ stemming from the fact that in the ($\bell, \x$) mixed representation $\bell$ plays the role of a constant of motion. Now, in contrast to what we found for $\b$ and $\K$,  there is no obvious classical equation of motion that would allow us to calculate directly the spread in $\x$, nor that in $\bell$. However, these dispersions of the off-diagonal elements (of a quantum nature) are controlled by the (classical) expectation values of $\K^2$ for $\langle \x^2\rangle$ and of $\b^2$ for $\langle \bell^2\rangle$.

The analysis  that has been presented in this section reveals that at late times, $t\gg t_2$, the density matrix becomes effectively diagonal both in momentum space and in coordinate space, reflecting the loss of quantum correlations associated to the off-diagonal elements. Note that this occurs sequentially, the suppression of the off-diagonal elements of the coordinate space density matrix occurring first, for time $t\gtrsim t_1$, while the  momentum space density matrix becomes diagonal only for $t\gtrsim t_2$. The fact that there seems to be no limit to this diagonalization is of course due to the absence of friction. We return to this issue at the end of  the next section.

As we have seen, the evolution of the density matrix is determined by three time scales
\begin{align}
t_0 = \frac{E}{\mu^2}\, ,\quad t_1=\frac{\mu^2}{\hat q} \, ,\quad t_2^3=\frac{E^2}{\hat q \mu^2}\, .
\end{align}
These time scales are not independent but are related to each other via the identity 
 \begin{align}
  t_2^3=t_1 t_0^2\, .
 \end{align}
Thus, there are only two possible time orderings: either $t_0>t_2>t_1$ or $t_0<t_2<t_1$. The former regime, with $t_0>t_1$,  appears to be the one of physical interest when discussing the dynamics of a highly energetic quark (see estimates below). In this regime, the collisions with the medium constituents dominate the time evolution and the spreading of the initial wave packet plays a minor role. This regime corresponds to an  initial wave packet well localized in momentum space, and correspondingly a broad density distribution. Formally this matches the small $\mu$, large $E$ limit, where the exact solution for the Wigner transform reads 
\begin{align}\label{strictEikonal}
 \rho_{_W}\!(\b,\K,t)_{E\to \infty}= \frac{4\mu^2}{\langle \k^2\rangle_t} e^{-\mu \b^2} e^{-\frac{\K^2}{\langle \K^2\rangle_t}}  \, ,
 \end{align}
 where $\langle \K^2\rangle_t=\mu^2+\hat q t\approx \hat q t $. In this limit, the motion of the quark in the transverse plane is negligible, and the sole effect of collisions is diffusion in momentum space.\footnote{The other time ordering, $t_0<t_2<t_1$, would correspond formally to the large $\mu$ limit. In that case the dominant phenomenon at early time would be the broadening of the density, amplified beyond $t_2$ by the collisions, while the effect of the collisions on the momentum distribution would not be  visible until $t\gtrsim t_1\gg t_2$. }

\begin{figure}[h!]
  \centering
  \includegraphics[width=.5\textwidth]{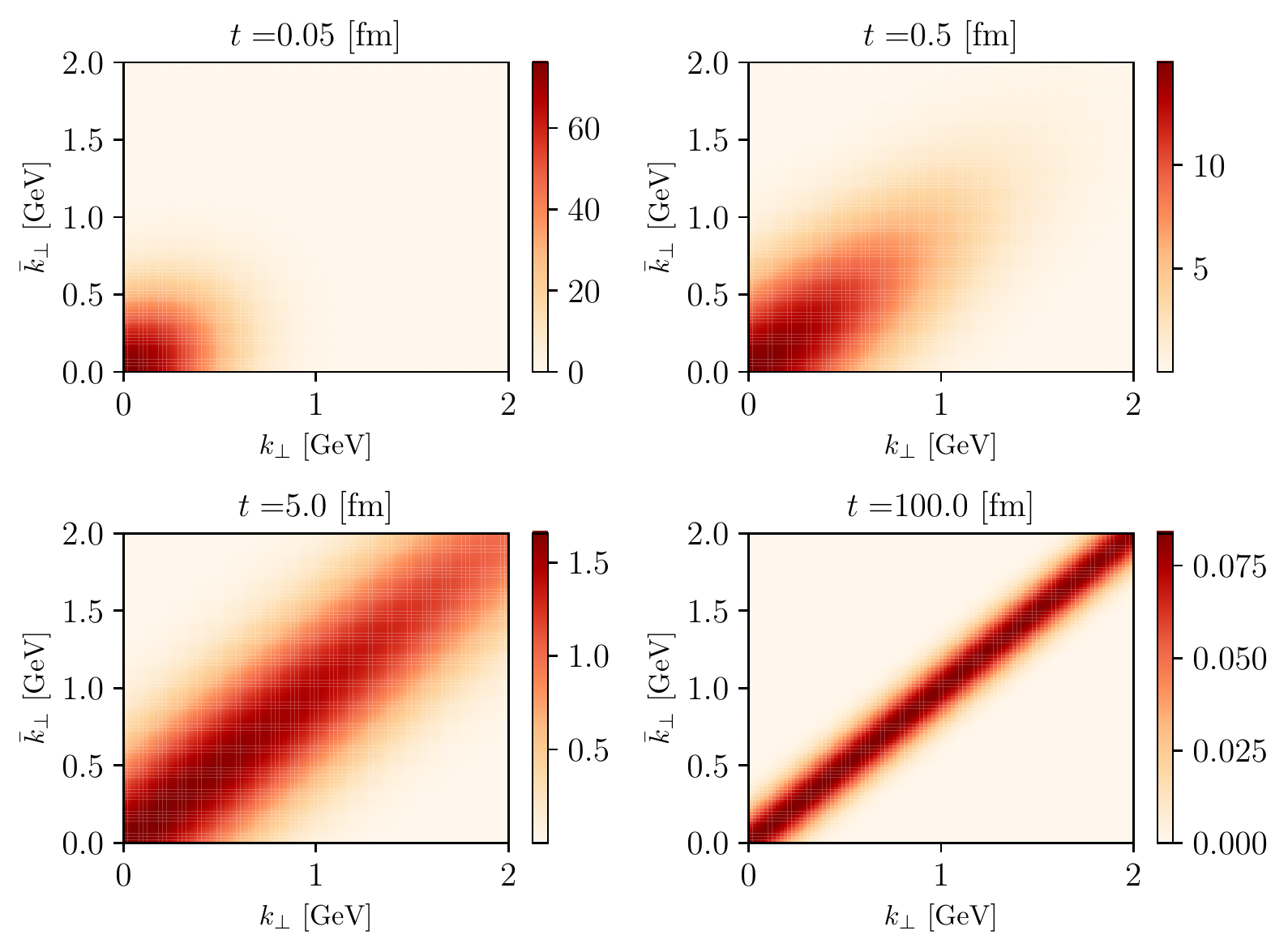}
  \includegraphics[width=.5\textwidth]{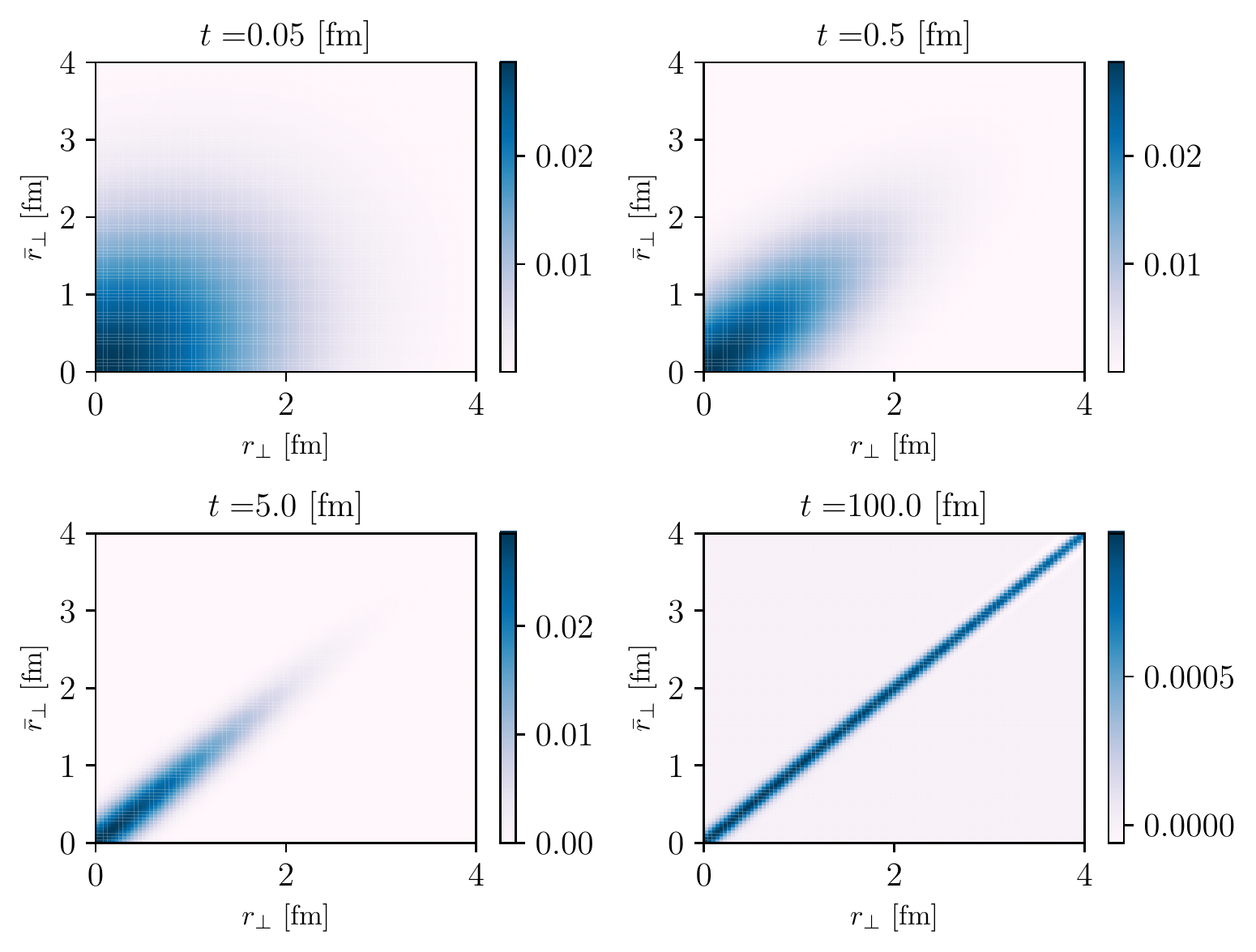}
  \caption{Time evolution of the real part of the quark density matrix, in momentum (top, red) and position (bottom, blue) space. We have used $\hat q=0.3 \, {\rm GeV}^3$, $\mu=0.3$ GeV, and $E=200 \, {\rm GeV}$. Also, $\k$ and $\bar \k$ (as well as $\r$ and $ \bar \r$) are taken to be parallel vectors. The first plot, in each representation, is in the region $t<t_1$, while the last plot is taken at very late times when $t>t_2$. The color axis is dimensionless.}
  \label{fig:plot_densitym} 
\end{figure}

To further illustrate all the points made so far, in Fig.~\ref{fig:plot_densitym} we plot the real part of $\bra{\k}\rho\ket{\bar \k}$ in terms of the absolute value of $\k$ ($k_\perp$) and $\bar \k $ ($\bar k_\perp$), for increasing evolution times and taking $\k \parallel \bar \k$.\footnote{See~\cite{refId0_HQ} for equivalent plots in the case of heavy quark evolution in the medium.} In addition, we provide the respective position space representation, $\bra{\r}\rho\ket{\bar \r}$ in terms of the absolute values of $\r$ ($r_\perp$) and $\bar \r $ ($\bar r_\perp$), with again $\r \parallel \bar \r$. For the chosen values of the parameters, $\hat q=0.3 \, {\rm GeV}^3$, $\mu=0.3$ GeV, and $E=200 \, {\rm GeV}$, we have $t_1\simeq 0.06$ fm, and $t_2\simeq 22.80 $ fm, while $t_0\simeq 444.44$ fm. The qualitative features discussed above are clearly visible. At early times, when $t<t_1$, the general pattern is set by the initial condition. As time increases, there is an overall decrease of the weight of each entry due to the underlying broadening of the distributions. At late time, when $t>t_2$,  the density matrix becomes diagonal. How this occurs depends on the representation. In coordinate space, the $\x$ dependence of $\rho(\b,\x)$ is controlled by momentum broadening, and the concentration along the diagonal becomes visible for $t\gtrsim t_1$. In momentum space, the same phenomenon occurs at later time, $t\gtrsim t_2$, and is controlled by the spreading of the quark density in the transverse plane. 

\begin{figure}[h!]
  \centering
  \includegraphics[width=.45\textwidth]{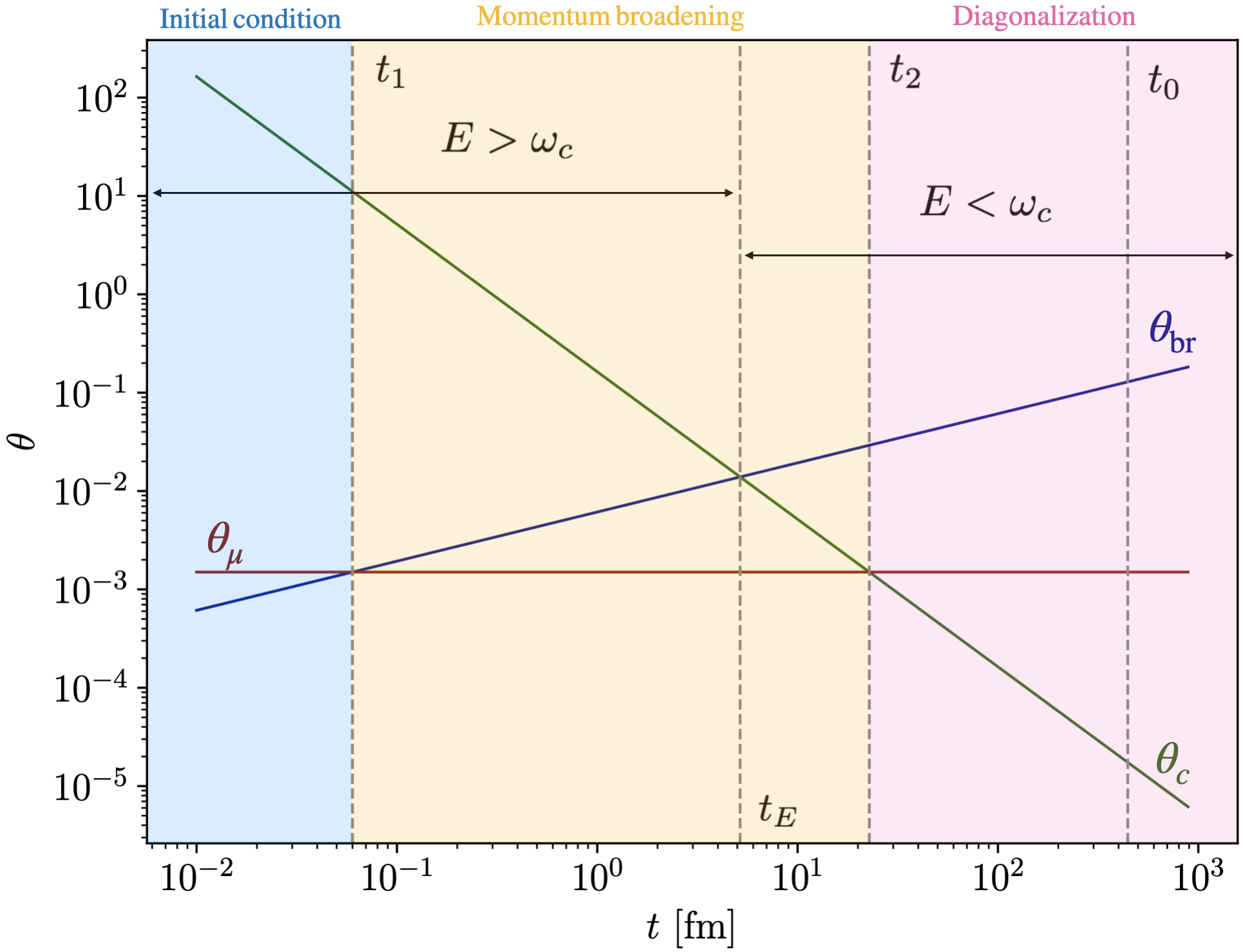}
  \caption{ Relation between the temporal and angular scales identified in the main text. We have identified three regions (shaded blue, yellow and pink regions), related to the evolution of $\rho$ in momentum space. In the first (blue), the initial condition sets the form of the density matrix. This is followed by a region (yellow) where momentum broadening drives diffusion in momentum space, but the off-diagonal elements are still sensitive to the initial configuration of the system. In position space, this region is characterized by the suppression of off-diagonal elements, while the diagonal has roughly the same form as the initial condition. Finally, at later times the system loses all knowledge of the initial condition (pink region). We used the values $\hat q=0.3 \, {\rm GeV}^3$, $\mu=0.3$ GeV, and $E=200 \, {\rm GeV}$. We also identify the scale when the quark energy matches the scale $\omega_c$, i.e. when $t=t_E$. This divides the plot into a region where $E>\omega_c$ and $E<\omega_c$.
  }
  \label{fig:scales}
\end{figure}

In view of making closer contact with the physics of jets, the initial motivation of this paper, it is  interesting to note that the previous discussion can be carried out in terms of dimensionless angular variables which are more closely related to jet observables. To that aim,  we introduce the following angles 
\begin{align}
  \theta_\mu^2 = \frac{\mu^2}{E^2} \, ,\quad \theta_c^2(t) =\frac{1}{\hat q t^3} \, ,\quad \theta^2_{\rm br}(t)= \frac{\hat q t}{E^2} \, .
\end{align}
Both $\theta_\mu$ and $\theta_{\rm br}$ have simple physical interpretations: $\theta_\mu$ measures the angle made by the quark momentum with respect to the initial direction of the high energy quark (the $z$-axis in Fig.~\ref{fig:doodle_draw}). If the initial distribution is very sharply peaked then $\theta_\mu\to 0$. The angle $\theta_{\rm br}$ has a similar interpretation, with the initial spread in transverse momentum substituted by the transverse momentum acquired by collisions. The angle $\theta_c$ depends only on $\hat q$ and time, but neither on $\mu$ nor $E$. Its role will be clarified shortly. Note that  
	$t_1$ is determined by $\theta_\mu=\theta_{\rm br}$ and $t_2$ by $\theta_\mu=\theta_c$, see Fig.~\ref{fig:scales}. At early times, i.e. when $t\ll t_1 \ll t_2 \ll t_0$,  the system is in the regime $\theta_{\rm br}\ll\theta_\mu\ll \theta_c$, and the dispersion around the diagonal of the density matrix in momentum space remains of order $\mu$, see Eq.~\eqref{eq:llop}. At later times, when $t> t_1$, one has $\theta_{\rm br}>\theta_\mu$, and, as time increases,  the average momentum becomes gradually dominated by the collisions with the medium constituents. Finally, when $t>t_2$, i.e., $\theta_c<\theta_\mu$, the off-diagonal elements of the momentum space density matrix decrease rapidly.  In this late time regime, the  momentum space density matrix takes the form (with $\theta_\ell\equiv |\k-\bar \k|/E$)
	 \begin{align}\label{rholKlate}
	 \rho(\bell,\K,t)\simeq \cP(\K,t) \exp\left\{ -\frac{1}{48}\frac{\theta_\ell^2}{ \theta_c^2}  \right\}\rme^{it \K\cdot\bell/(2E)}\, ,
  \end{align}
	 and gradually identifies to the momentum distribution as $t$ grows. Fig.~\ref{fig:scales} also reveals the existence of a  cross-over  time  $t_E$ where $\theta_c=\theta_{\rm br}$:
\begin{align}
 t_E \equiv \sqrt{\frac{E}{\hat q}}=\sqrt{t_0t_1}, \qquad t_1< t_E<t_2.
\end{align}
This corresponds to the time where the energy of the quark satisfies $E= \hat q t^2 \equiv \omega_c$. However, neither the time scale $t_E$ nor the energy $\omega_c$  seem to  play any significant role in the evolution of the density matrix.

Interestingly, the angle $\theta_c$ has been previously identified in the context of color coherence of QCD antennas in the medium~\cite{Casalderrey-Solana:2012evi,Caucal:2018dla,Barata:2021byj,Dominguez:2019ges,Casalderrey-Solana:2011ule,Mehtar-Tani:2012mfa}. It is also related to the characteristic angle for medium induced radiation~\cite{Zakharov:1996fv,Baier:1996sk}. In the present case, $\theta_c$ controls the loss of coherence between the two legs of the effective dipole in the third diagram in Fig.~\ref{fig:master_draw}. As discussed earlier, this loss of coherence, that manifests itself in the  vanishing of the off-diagonal matrix elements of $\bra{\k}\rho\ket{\bar \k}$, can be related to the random walk done by the quark in the transverse plane. The scale $\omega_c$ is also associated to the  medium induced radiation problem, characterizing the critical frequency above which gluon emissions are suppressed due to the QCD LPM effect \cite{Baier:1996sk,Zakharov:1996fv} and is  related to  $\theta_c$ by $\theta_c = \sqrt{\langle \k^2\rangle_t}/\omega_c$.

\section{Entropy as a measure of quantum to classical transition}
\label{sec:entropy}
The two regimes that we have identified in the previous section, with their associated characteristic time scales $t_1$ and $t_2$, are nicely reflected in the entropy associated with the density matrix. There are several entropy measures which, although differing quantitatively, are maximized/minimized by the same density matrices, and share common qualitative features, see e.g.~\cite{Witten:2018zva} and references therein. Here we shall consider the von Neumann entropy which we compare to the classical entropy associated with the Wigner representation of the density matrix. We shall also mention the Renyi-2 entropy which, in the present context, carries essentially the same information as the von Neumann entropy. 

The von Neumann entropy $S_{\rm vN}$ is defined as 
\begin{align}\label{eq:vN-entropy-def}
 S_{\rm vN}[\rho] =-  \Tr  \rho \log   \rho \, .
\end{align}
In the present case, the calculation of $S_{\rm vN}$ reduces to the calculation of multidimensional Gaussian integrals. It is explicitly carried out in Appendix~\ref{se:vNentropy}.   It is shown there that  $S_{\rm vN}$ depends on time through a single variable, which can be chosen to be the so-called purity $p$
\begin{align}\label{eq:purity}
p\equiv \Tr \rho^2= \frac{E^2}{a c -\frac{b^2}{4}}.
\end{align}
The purity measures the deviation of the density matrix from that of a pure state. 
For a pure state, $\Tr \rho^2=\Tr \rho=1$, and $p=1$. When the density matrix represent a statistical mixture, $p<1$. 
In terms of $p$, $S_{\rm vN}$ takes the form
\begin{align}\label{eq:S1}
 S_{\rm vN} 
 &=  \log \left(\frac{1-p}{4p}\right) + \frac{1}{\sqrt{p}}\,  \log  \frac{1+p +2 \sqrt{p}}{(1-p)}\,.
\end{align}
By using the explicit time dependence of the coefficients $a$, $b$ and $c$ given by \eqn{eq:abc}, one obtains
\begin{align}\label{eq:purity-2}
\frac{1}{p}=\left(1+\frac{ t}{t_1}\right) \left( 1 + \frac{t^3}{12t_2^3}\,\frac{t+4t_1}{ t+t_1}\right)  .
\end{align}
As discussed in the previous section, the following ordering of time scales holds in the relevant high energy regime: $t_1 < t_2 <t_0$. There are therefore three regimes of interest :
\begin{itemize}
\item Initial stage: 
\beq
t \ll  t_1,\qquad 
p \simeq  1  \,.
\eeq
In this stage, collisions do not play a significant role and the system evolves as  a nearly pure state with vanishing entropy,  $S_{\rm vN} \to 0$ as $p\to1$.

\item Spatial decoherence due to collisions:  $t_1 \ll  t \ll  t_2$
\begin{align}\label{eq:lllplp}
p \simeq  \frac{1}{\left(\frac{ t}{t_1}\right) \left( 1 + \frac{t^3}{12t_2^3}\right)}\simeq  \frac{t_1}{t}  \ll 1\,.
\end{align}
In this regime, $p\ll 1$ and the system is in a mixed state. Moreover, the off-diagonal elements of the density matrix in the position representation become strongly suppressed by the growth of  $\langle \K^2\rangle_t$ (see Eq.~(\ref{rhob0x2})). 
In this regime, the von Neumann entropy takes the simple form: 
\begin{align}\label{eq:vN-limit-p}
S_{\rm vN}&  = \log  \frac{1}{p}+2-\log 4+\cO(\sqrt{p}) \,\, \nn
 & \underset{t_1\ll t\ll t_2 }{\simeq}  \log  \frac{t}{t_1}=\log \frac{\langle \K^2\rangle_t}{ \mu^2}  \,.
\end{align}
It increases logarithmically with  $\langle \K^2\rangle_t$. Note the constant, $2-\log 4$, beyond the leading logarithmic term. We shall encounter it again shortly.

\item Full diagonalization and memory loss: $t\gg  t_2 $
\begin{align}\label{eq:p-limit-3}
p \simeq  \frac{12 t_1 t_2^3 }{t^4}\ll 1\,.
\end{align} 
What characterizes this regime is the diagonalization of the density matrix in the momentum space, as well as  the loss of memory of the initial condition. Asymptotically the entropy behaves as 
\begin{align}\label{eq:final_S1}
  S_{\rm vN}   &\simeq \log  \frac{1}{p} \simeq \log   \frac{\hat q^2 t^4}{E^2} \sim \log  \langle \k^2\rangle_t  \langle \b^2\rangle_t \, .
\end{align}
Its increase as the logarithm of the  total phase space measured by $\langle \k^2\rangle_t  \langle \b^2\rangle_t$ is what one could expect in a classical regime.
\end{itemize}

To confirm this classical feature,  let us consider the following  Wigner entropy 
\begin{align}\label{eq:W-entropy}
S_{_{\rm W}} \equiv - \int_{\K,\b} \rho_{_{\rm W}} (\b,\K) \log  \rho_{_{\rm W} }(\b,\K) \, ,  
\end{align}
where $\rho_{\rm W} $ is the Wigner representation of the density matrix, see e.g.~\cite{Hagiwara:2017uaz,VanHerstraeten:2021nce}. 
\eqn{eq:W-entropy} can be understood as the classical entropy obtained from the phase space distribution associated with $\rho_{_W}$. A simple calculation yields
 \begin{align}\label{eq:W-entropy-p}
S_{_{\rm W}}&  = \log  \frac{1}{p}+2-\log 4 .
\end{align}
The constant $S_{_{\rm W}}(t=0)=2-\log  4$, is that met earlier in the small $p$ expansion of $S_{\rm vN}$ (see Eq.~(\ref{eq:vN-limit-p})). It  corresponds to the entropy of a classical distribution function that takes the same form as the initial density matrix. Comparing \eqn{eq:W-entropy-p} to \eqn{eq:vN-limit-p} we find that at late times
\begin{align}\label{eq:W-entropy-p_2}
\frac{S_{_{\rm W}}-S_{\rm vN}} {S_{_{\rm W}}}&  \approx \frac{\sqrt{p}}{\log (1/p)} . 
\end{align}
This vanishes as $ 1/(t^2 \log t)$ when $t\to\infty$, indicating that  at late times the entropy content of the quark density matrix coincides with that of  a classical phase space distribution. 

The Renyi-2 entropy is often used in place of the von Neumann entropy, because it is much simpler to evaluate. It is defined as~\cite{nielsen_chuang_2010}
\begin{align}\label{eq:S_alpha}
S_2[\rho]= -\frac{1}{2} \log (\Tr \, \rho^2) = \frac{1}{2}\log\frac{1}{ p}\, ,
\end{align} 
Thus, to within a factor $1/2$ and the constant $2-\log 4 $, it is identical to the Wigner entropy introduced in Eq.~(\ref{eq:W-entropy-p}). At late time all three entropies, the von Neumann entropy, the Renyi-2 entropy and the Wigner entropy carry the same information. 

\begin{figure}
  \centering
  \includegraphics[width=.48\textwidth]{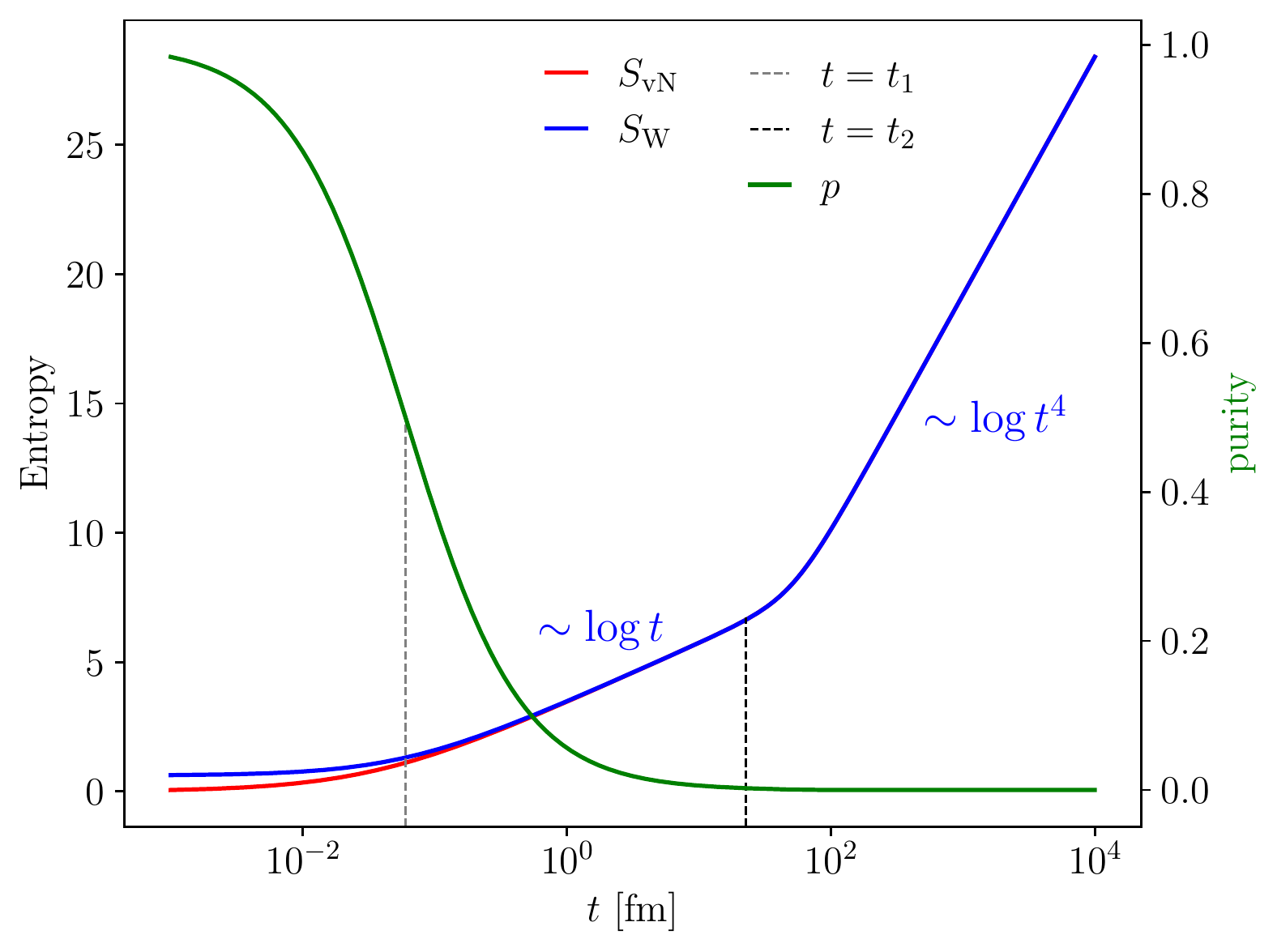}
  \caption{Time evolution of the quark's $S_{vN}$ and $S_W$ entropies. Here we used the same parameters as in Fig.~\ref{fig:plot_densitym}.}
  \label{fig:entropy}
\end{figure}

To illustrate all the above points,  we display in Fig.~\ref{fig:entropy} the time evolutions of both the von Neumann and the Wigner entropies. This figure confirms  that $S_{\rm vN}\simeq S_{_{\rm W}}$ as soon as $t\gtrsim t_1$.  In addition, the figure displays  the drop of the purity  when  $t\sim t_1$.

As a final remark,  we note that the apparent unbounded growth of the entropy at late times, is a consequence of the fact that the master equation for the reduced density matrix that we have used accounts only for collisional decoherence, but ignores friction and dissipation. We could expect indeed friction to alter the late time evolution and drive the transverse motion of the quark to equilibrium with the surrounding medium. It is relatively straightforward to include such effects in the master equation, following for instance the analogy with the heavy quark problem  \cite{Blaizot:2017ypk}. Including friction effect  amounts in particular to generalize the correlation function in Eq.~(\ref{eq:2-pt-correlator}) beyond the instantaneous limit. 
In momentum space, this leads generically to an evolution equation of the form\footnote{We postpone the discussion of the complete evolution equation for the density matrix to a forthcoming work.} (cp. Eq.~(\ref{diffisionPK}))
\begin{align}\label{eq:Boltzmann}
\left[ \frac{\partial}{\partial t} -   \frac{\partial}{\partial \K} \left(\frac{\hat q}{4} \frac{\partial}{\partial \K}+ \gamma_f \frac{\K}{E}\right)\right]\cP(\K,t)= 0\, , 
\end{align}
where we used the harmonic approximation and $\gamma_f$ is the friction coefficient, related to $\hat q$ via the Einstein relation,  $\gamma_f=\hat q / 4T\sim T^2$. The late time solution to this equation is a thermal momentum distribution, $\cP(\K,t)\propto \exp[-\K^2/(2 E  T)]$. This implies that at late times the entropy saturates at a maximum value, which is achieved once $\langle \K^2 \rangle_t\simeq\hat q t = 2 E T$,  that is at time  $t\gtrsim t_{\rm rel}$, where $ t_{\rm rel} \equiv ET/\hat q$. When this transverse thermalization is reached,  the off-diagonal matrix elements of the coordinate space density matrix remain significant in a small range of the order of the thermal wavelength $\lambda_T\sim 1/ET$, rather than becoming arbitrarily small. For  the present choice of parameters, $\hat q=0.3 \, {\rm GeV}^3$, $T=0.5$ GeV, and $E=200 \, {\rm GeV}$,  we find $t_{\rm rel}\simeq 66.7$ fm, which is larger than $t_2\simeq 22$ fm,  but not much larger.\footnote{Note that $t_{\rm rel}=t_E^2 T$ is independent of the initial condition, but $t_2$ is.}  Thus, whether, in the presence of dissipation, the regime of the rapid entropy growth will develop fully  or not may depend significantly on the values of the relevant  parameters.

\section{Summary and conclusions}
\label{sec:sum}

We have studied the time evolution of the density matrix of an energetic quark in the presence of a dense QCD medium, by solving simple master equations for its reduced density matrix. The  two  color components of the density matrix evolve independently, with only the singlet component surviving the propagation through the QCD medium. At late times we observe that the off-diagonal elements  are suppressed sooner in the position representation than  in the momentum representation of the density matrix. We interpret the simultaneous diagonalization of the density matrix in both representations as a signal that its Wigner representation behaves at late times as a classical phase space distribution. This interpretation is further supported by the explicit calculation of the von Neumann entropy, which is shown to agree completely with the classical Wigner entropy at late time. The entropy growth can be split in two regimes where first the momentum space is populated, followed by position space. We note that the regime of the fast entropy growth at late time  is not accessible in the exact eikonal limit, i.e. $E\to \infty$, where the density matrix reduces to Eq.~(\ref{strictEikonal}), because in this regime the transverse position of the quark is frozen. We also argue that this regime may be affected by dissipation effects that are not included in the master equation that we used. A simple estimate of these dissipation effects suggests that the development of the late time entropy growth may be tamed before developing fully, depending on the values of the relevant parameters.

The model that we have studied in this paper is an oversimplified picture of an in-medium  jet.  More realistic situations can of course be considered, such as more complex initial conditions for the quark, and a more realistic description of the plasma.  More importantly the present study needs to be extended by including radiative corrections to the reduced density matrix. Such radiative correction would, in particular,  have impact on the jet entropy  \cite{Neill:2018uqw}.  In addition, increasing the number of particles in the jet leads to more complex color structures that are worth exploring~\cite{Blaizot:2012fh,Apolinario:2014csa}. Although the in-medium evolution results in the randomization of the jet's color degrees of freedom~\cite{Zakharov:2018hfz}, how such an effect manifests itself beyond two particles in the final state is not known in detail so far. Understanding this aspect is important to describe the final particle distribution after hadronization, see e.g.~\cite{Beraudo:2011bh}. Such issues will be addressed in future works.

\section*{Acknowledgements}
J. B. is grateful to Andrey Sadofyev and Xin-Nian Wang for helpful discussions. Y. M.-T. and J. B.'s work has been supported by the U.S. Department of Energy under Contract No.~DE-SC0012704. Y. M.-T. acknowledges support from the RHIC Physics Fellow Program of the RIKEN BNL Research Center.  J.-P. B. acknowledges partial support from the
European Union's Horizon 2020 research and innovation
program under grant agreement No 824093 (STRONG-
2020).\\

\appendix

\section{Derivation of the color singlet and octet evolution equations}\label{app:dev_13_14}

Here we derive the master equations for the singlet and octet color components of the density matrix, i.e. Eqs.~\eqref{eq:rho_s} and \eqref{eq:rho_o}. For that, we start from the relation which connects the density matrix at different times:
\begin{align}\label{eq:111}
  \bra{\k}\rho^{ij}(t+\delta t )\ket{\bar \k}=\int_{ \q ,\bar \q} &\bra{\q} \rho^{i_0j_0}(t) \ket{\bar \q} \nn 
  &\hspace{-3.5 cm}\times \langle \cG_{i i_0}(\k,t+\delta t;\q,t) \cG^\dagger_{ j_0 j}(\bar \k,t+\delta t;\bar \q,t)\rangle \, ,
\end{align}
where $\delta t$ should be understood as being a small time step, see Fig.~\ref{fig:master_draw}. The propagator $\cG$ describes the evolution of a single particle in the presence of a classical background. It is the Green's function to Eq.~\eqref{SchroG} and its explicit path integral representation can be found e.g.~\cite{Blaizot:2012fh,Blaizot:2015lma} and references there in. Importantly, it satisfies the relation (in coordinate free form)
\begin{align}
  \cG(t;u) =  \cG_0(t;u) -ig \int_{s} \cG_0(t;s)A^-(s) \cG(s;u) \, ,
\end{align}
with $u<s<t$. Using the explicit form of the vacuum propagator
\begin{align}\label{eq:G0_k}
    \cG_0(\k,t;\q,s)=  (2\pi)^2 \delta(\k-\q) e^{-i \frac{\k^2}{2E}(t-s)}  \, ,
\end{align}
we can derive that
\begin{widetext}
  \begin{align}\label{eq:help1}
    \cG(\k,t+\delta t;\q, t) &\cG^\dagger(\bar \k,t+\delta t;\bar \q,t) =(2\pi)^2  e^{-i \frac{(\k^2-\bar \k^2)}{2E}\delta t} \Bigg[ (2\pi)^2 \delta(\k-\q)\delta(\bar \k-\bar \q)\nn
    &+\frac{g^2}{(2\pi)^2}\int_{t}^{t+\delta t} ds_1 \int_{t}^{t+\delta t} ds_2 \, e^{i\frac{\k^2}{2E}s_1}A(\k-\q,s_1) e^{-i\frac{\q^2}{2E}s_1 } \, e^{-i\frac{\bar \k^2}{2E}s_2}A^\dagger(\bar \k-\bar \q,s_2) e^{i\frac{\bar \q^2}{2E}s_2 }\nn
    &-g^2  \delta(\bar \k-\bar \q) \int_{\p} \int_t^{t+\delta t} ds_1 \int_t^{s_1} ds_2\, e^{i\frac{\k^2}{2E}s_1}A(\k-\p,s_1) e^{-i\frac{\p^2}{2E}(s_1-s_2)}A(\p-\q,s_2) e^{-i\frac{\q^2}{2E}s_2}\nn
    &-g^2  \delta(\k-\q) \int_{\p}\int_t^{t+\delta t} ds_1 \int_t^{s_1} ds_2 \, e^{-i\frac{\bar \k^2}{2E}s_1}A^\dagger(\bar \k-\p,s_1) e^{i\frac{\p^2}{2E}(s_1-s_2)}A^\dagger(\p-\bar \q,s_2) e^{i\frac{\bar \q^2}{2E}s_2}  \Bigg]\nn 
    & +\mathcal{O}((gA)^3)\, .
  \end{align}  
\end{widetext}
Here we already dropped the terms linear in the background field since they vanish after tracing out the medium according to Eq.~\eqref{eq:2-pt-correlator}. We also used the shorthand notation $A=A^{a,-}t^a$, and we have left the color indices implicit. Notice that the contributions are ordered according to the diagrams in Fig.~\eqref{fig:master_draw}.

To obtain the evolution equations from Eq.~\eqref{eq:help1}, we average over the background field and expand to leading order in $\delta t$. Since the averaging according to Eq.~\eqref{eq:2-pt-correlator} is local in time, one can directly construct the associated integral evolution kernel.\footnote{The locality of the medium averages will remove some of the quantum effects~\cite{Blaizot:2018oev,Blaizot:2017ypk} associated to the quark's evolution in the medium. Such non-local contributions are, for example, responsible for the transverse space thermalization discussed in the main text.} Also since the average has Gaussian statistics, there is no need to include higher order terms in the background field of Eq.~\eqref{eq:help1}.  It is straightforward to show that the real single gluon exchange, given by the second line in Eq.~\eqref{eq:help1}, generates the contribution to the density matrix at time $t$
\begin{align}
  \int_{ \q } \int_0^t dt' &e^{-i \frac{(\k^2-\bar \k^2)}{2E}(t-t')}  \gamma(\q) \nn 
  &\hspace{0 cm}\times  t^a_{ii_0}t^a_{j_0j}\bra{\k-\q} \rho^{i_0j_0}(t')  \ket{\bar \k-\q}\, ,
\end{align}
while the virtual terms (last two lines in Eq.~\eqref{eq:help1}) yield 
\begin{align}
  -\frac{1}{2} &\int_{ \q } \gamma(\q)  \int_0^t dt' e^{-i \frac{(\k^2-\bar \k^2)}{2E}(t-t')}\,  \nn 
  &\times \Bigg\{   t^a_{ik}t^a_{ki_0}  \delta_{j,j_0}+ t^a_{j_0k}t^a_{kj} \delta_{i,i_0}\Bigg\}   \bra{\k} \rho^{i_0j_0}(t')  \ket{\bar \k}   \, .
\end{align}
Notice that here we have promoted the small time step $\delta t$ to an integral from the initial time $t=0$ up to time $t$, using that the problem is invariant under time translations. The singlet and octet decomposition of these contributions is obtained from applying Eq.~\eqref{eq:os_decomp}.

Focusing on the singlet case, after absorbing the trivial contribution associated to zero gluon exchanges, we find that the above terms lead to the following relation
\begin{align}\label{eq:iii_app}
  \langle \k | \rho_{\rm s}(t)| \bar \k\rangle&=C_F  \int_{\q}  \int_0^t dt' \, e^{i\frac{(\k^2-\bar \k^2)}{2E}(t-t')} \nn 
   &\times  \gamma(\q)\left[   \langle \k-\q | \rho_{\rm s}(t')| \bar \k-\q\rangle     -  \langle \k | \rho_{\rm s}(t')| \bar \k\rangle   \right]\, .
\end{align}
In the coordinates
\begin{align}
\K= \frac{\k+\bar \k}{2}\,,  \quad \bell= \k-\bar \k\,  , 
\end{align}
the action of the operator 
\begin{align}\label{eq:klop}
\rme^{i\frac{\K\cdot \bell}{E} t} \frac{\del }{\del t}\rme^{-i\frac{\K\cdot \bell}{E} t} =   \frac{\del }{\del t} -i\frac{\K\cdot \bell}{E} \,
\end{align}
on Eq.~\eqref{eq:iii_app} yields the momentum space master equation for the singlet
\begin{align}
 &\frac{\del }{\del t}  \langle \K+\bell/2 | \rho_{\rm s}(t)| \K-\bell/2\rangle= \nn 
 &-i\frac{\K\cdot \bell}{E} \langle \K+\bell/2 | \rho_{\rm s}(t)|  \K-\bell/2\rangle    \nn
 &+  C_F \int_{\q} \gamma(\q)[   \langle \K+\bell/2-\q | \rho_{\rm s}(t)| \K-\bell/2-\q\rangle    \nn 
 &-  \langle \K+\bell/2 | \rho_{\rm s}(t)|  \K-\bell/2\rangle  ]\,.
\end{align} 
After transforming to the $(\bell,\x)$ mixed coordinate representation, this reduces to Eq.~\eqref{eq:rho_s}.

\section{Calculation of the von Neumann entropy}
\label{se:vNentropy}

The calculation of the von Neumann entropy can be efficiently done by applying the replica trick:
\begin{align}\label{eq:vN-replica}
 \Tr  \rho \log   \rho =  \frac{\del}{\del n }\Tr  \rho^n \Big|_{n=1}  \, .
\end{align}
To compute $\Tr\rho^n$, we note that, in  momentum space, 
\begin{align}
\langle \q_2 | \rho| \q_1 \rangle  = \frac{4\pi}{a} \exp\left[ -\frac{1}{2a } \left( \alpha(\q_2^2+\q_1^2) -  \beta \q_1\cdot \q_2 \right)\right] \, ,
\end{align}
where 
\begin{align}\label{eq:beta}
 \beta &= \frac{1}{2E^2} \left(a c-\frac{b^2}{4}\right) -\frac{1}{2}, \,\alpha=\beta+1\,,
\end{align}
and we have used Eqs.~\eqref{eq:abc} and related definitions.  We then write the trace of the $n$th power of $\rho$ as 
\begin{align}\label{eq:aaaa}
 \Tr \rho^n &= \langle \q_{{n+1}}|\rho|\q_n\rangle \langle \q_{{n}}|...|\q_2\rangle \langle \q_{{2}}|\rho|\q_1\rangle \nn
 &= \left(\frac{4\pi}{a}\right)^n \int_{\q_1,\q_2,\cdots \q_n} e^{-\frac{1}{a}\Q^T\cdot  \M  \cdot\Q} \nn 
 &= \frac{1}{\det(\M)}\, ,
\end{align}
with the periodic boundary conditions $\q_{n+1}=\q_1$. Here $\Q$ is a column vector whose components are the $n$ momenta $\q_1$ to $\q_n$. The matrix $\M$ is a $n\times n$  matrix that we shall specify shortly.

First, we should discuss the special cases $n=1$ and $n=2$ for which the matrix $\M$ does not obey the general form valid for $n>2$. The case $n=1$  is straightforward since $\M\equiv  \alpha+\beta=1$. Hence, $\det(\M)=1$ and we simply recover that $\Tr \rho=1$. 
The $n=2$ case corresponds to the  purity~\cite{nielsen_chuang_2010}
\begin{align}
p\equiv \Tr \rho^2= \int_{{\q_1,\q_2}} \langle \q_2 | \rho| \q_1 \rangle^2 \,.
\end{align}
The matrix $\M$ in \eqn{eq:aaaa} is then
\begin{align}
 \M =\begin{pmatrix}
\alpha & -\beta \\
-\beta & \alpha 
\end{pmatrix}\,, 
\end{align}
which yields 
\begin{align}\label{eq:purity-4}
p&= \frac{1}{\alpha^2-\beta^2}= \frac{1}{2\beta+1}=\frac{E^2}{a c -\frac{b^2}{4}} 
\, .
\end{align}

In the case $n>2$, the  matrix $\M$ reads
\begin{align}
 \M \equiv  \begin{pmatrix}
 \alpha & -\beta/2 & 0 & \cdots& 0  & -\beta/2 \\
 -\beta/2  &  \alpha &-\beta/2& \cdots & \cdots   & 0 \\
 0& -\beta/2 & \alpha & -\beta/2  &\cdots & 0 \\
\vdots  &  \ddots & \ddots &  \ddots  & \ddots& \vdots  \\
\vdots  & \vdots  & \ddots    &  \ddots &  \ddots &-\beta/2   \\
 -\beta/2 & 0 & \cdots & 0  &-\beta/2  &\alpha
 \end{pmatrix}\, ,
\end{align}
 Besides the top right and bottom left elements $M_{n,1}=M_{1,n}=-\beta/2$, that reflect the boundary conditions $\q_{n+1}=\q_1$,  $\M$ is a tridiagonal symmetric Toeplitz matrix, with $M_{i,i} =\alpha$, $M_{i,(i+1)} =-\beta/2$, $M_{(i+1),i} =-\beta/2$ and $M_{i,j} =0$ otherwise. To compute the determinant of $\M$ we use the fact that the determinant of a tridiagonal matrix $\T (x)$, with off-diagonals equal to $1$ and the diagonal elements set to $2x$, satisfies 
\begin{align}
 \det (\T(x)) = U_n(x)\, , 
\end{align}
where $U_n(x)$ is the Chebyshev polynomial of the second kind, which can be defined through
\begin{align}
 U_n(\cos(\theta)) = \frac{\sin((n+1)\theta)}{\sin(\theta)} \, .
\end{align}
Using these elements, one can show that 
\begin{align}
 \det (\M) =  \left(\frac{\beta}{2}\right)^n \left[U_n(x) - U_{n-2}(x) + 2 \right] \, ,
\end{align}
where $x=\alpha/\beta$. By using  the identity 
\begin{align}
U_n(x) -  U_{n-2}(x) &= \left(x-\sqrt{x^2-1}\right)^n \nn 
&+\left(x+\sqrt{x^2-1}\right)^n  \, ,
\end{align}
we finally obtain
\begin{align}\label{eq:rho_n}
& \Tr\rho^n = \left(\frac{2}{\beta}\right)^n \, \frac{1}{\left(x-\sqrt{x^2-1}\right)^n +\left(x+\sqrt{x^2-1}\right)^n  +2}\,.
\end{align}
By combining this result with \eqn{eq:vN-replica}, we get  the von Neumann entropy  in the following form
\begin{align}\label{eq:S1a}
 S_{\rm vN} &=  \log \left(\frac{\beta}{2}\right) + \frac{1}{2} \sqrt{x+1} \log  \frac{x+\sqrt{x^2-1}}{x-\sqrt{x^2-1}}\nn
 &=  \log \left(\frac{\beta}{2}\right) + \sqrt{2\beta+1}\,  \log  \frac{\beta+1+\sqrt{2\beta+1}}{\beta}\,.
\end{align}
The expression (\ref{eq:S1}) of the main text is obtained by expressing $\beta $ in terms of $p$ using Eq.~(\ref{eq:purity}).

\bibliographystyle{apsrev4-1}
\bibliography{references.bib}

\end{document}